\documentclass[a4paper,11pt]{article}

\usepackage{jheppub} 

\usepackage[T1]{fontenc} 
\usepackage{braket}
\usepackage{csquotes}
\usepackage{nameref}
\usepackage{amsmath}
\usepackage{xspace}
\usepackage{tikz}
\usetikzlibrary{mindmap,chains}
\usetikzlibrary{calc}

\def\bra#1{\left\langle#1\right|}
\def\ket#1{\left|#1\right\rangle}

\def\kc#1{\left(#1\right)}
\def\kd#1{\left[#1\right]}

\def\Im{\mathrm{Im}}

\def\=>{\Rightarrow}
\def\>{\rightarrow}

\def\Tr{\mathrm{Tr}}

\def\R{\mathcal{R}}

\title{\boldmath Symmetry-Resolved Entanglement Entropy from Heat Kernels}

\author[a]{Yuan-Chun Jing}
\author[a]{Chao Niu$^{*}$}
\author[b]{Zhuo-Yu Xian$^{*}$}

\affiliation[a]{Department of Physics, Jinan University, Guangzhou 510632, China}
\affiliation[b]{Department of Physics, Freie Universit\"at Berlin, Arnimallee 14, DE-14195 Berlin, Germany}

\emailAdd{jingyuanchun18@mails.ucas.ac.cn}
\emailAdd{niuchaophy@gmail.com}
\emailAdd{zhuo-yu.xian@fu-berlin.de}

\footnotetext{Corresponding author.}

\abstract{We develop a systematic framework for computing symmetry-resolved entanglement entropies (SREE) in charged quantum systems based on an improved heat kernel approach. 
Although the conventional Sommerfeld formula proves effective for neutral systems, it encounters limitations when gauge fields or chemical potentials are introduced due to incomplete residue prescriptions and violations of asymptotic boundary conditions. 
By reconstructing the analytic structure of the heat kernel using a phase factor, we derive a globally convergent expansion that reconciles discrete residue summations with continuous spectral decompositions. 
We further apply this framework to Gaussian continuous multi-scale entanglement renormalization ansatz (cMERA) states and show that the entanglement entropy (EE) can be expressed in terms of the cMERA flow functions. In particular, we obtain a symmetry-resolved entanglement entropy flow equation in the presence of a chemical potential.
This formulation extends naturally to arbitrary spacetime dimensions and recovers established results for neutral systems in the $\mu\to0$ limit. 
We validate our framework through two settings: (1) exact agreement with $(1+1)$-dimensional conformal field theory (CFT) predictions using twist-operator techniques, and (2) consistency with holographic entropy calculations on $S^1\times H^{d-1}$ geometries. 
Our results both unify the treatment of charged and neutral entanglement entropy and extend this treatment to real-space renormalization frameworks, providing a robust tool for probing symmetry-resolved entanglement in conformal field theories, their holographic duals, and cMERA representations.}

\begin{document}
\maketitle
\flushbottom

\section{Introduction}\label{sec:intro}
Entanglement entropy (EE) has become a cornerstone for understanding emergent phenomena in quantum many-body systems, from critical spin chains to holographic spacetimes. 
At its core, entanglement entropy quantifies the irreducible correlations between subsystems, a measure that has proven indispensable in characterizing phases of matter, quantum chaos, and the very fabric of spacetime within AdS/CFT correspondence \cite{Eisert:2008ur, Eisert:2006kue, Horodecki:2009zz, Levin:2006zz, Holzhey:1994we,Calabrese:2009qy, Calabrese:2004eu,Calabrese:2016xau,Calabrese:2009kka,Cardy:2007mb,Casini:2005rm,Casini:2005zv,Casini:2009sr,Casini:2011kv,Ryu:2006bv,Miyaji:2015yva,Swingle:2009bg,Vidal:2007hda,Vidal:2008zz,Hung:2011nu,Balasubramanian:2011wt,Balasubramanian:2013lsa,Balasubramanian:2013rqa,Amico:2007ag,Nishioka:2018khk,Latorre:2009zz,Lewkowycz:2013nqa}. However, when one moves beyond the neutral sector toward symmetry-resolved entanglement \cite{Belin:2013uta,Goldstein:2017bua,Xavier:2018kqb}, fundamental questions naturally arise: How do conserved charges imprint their structure onto entanglement and what mathematical tools can universally capture this interplay between symmetry and quantum correlations?

The interplay between quantum entanglement and global symmetries has emerged as a unifying framework across diverse physical domains, from holographic quantum gravity and topological phases to critical many-body systems.
This synergy, driven by novel theoretical constructs \cite{Belin:2013uta,Belin:2014mva,Pastras:2014oka,Matsuura:2016qqu,Laflorencie:2014cem,Xavier:2018kqb,Goldstein:2017bua}, reveals profound connections between information-theoretic principles and collective quantum phenomena. Importantly, the decomposition of entanglement entropy into distinct symmetry sectors,namely symmetry-resolved entanglement entropy (SREE), goes beyond mere theoretical elegance. It encodes the fine-grained quantum correlations constrained by symmetry, thereby offering insights that are inaccessible through total entanglement measures alone. Remarkably, the equipartition, namely,
SREE becomes approximately independent of the symmetry sector at leading order, stands as a hallmark property of SREE in gapless systems. 

The foundation of SREE lies in quantum information theory, where the Holevo theorem rigorously quantifies the accessible information in symmetry-decomposed mixed states: the maximal extractable information is governed by the entropy difference, which reduces to Shannon entropy in specific limits \cite{Xavier:2018kqb}. Furthermore, exact computations across diverse theories have established SREE as a universal probe \cite{Goldstein:2017bua,Xavier:2018kqb,Capizzi:2020jed,Calabrese:2021wvi,Bonsignori:2020laa,Estienne:2020txv,Ma:2021zgf,Ares:2022gjb,Ghasemi:2022jxg,Murciano:2020vgh,Horvath:2020vzs,Horvath:2021fks,Horvath:2021rjd,Capizzi:2021kys,Capizzi:2022jpx,Capizzi:2022nel,Foligno:2022ltu,Fraenkel:2019ykl,Fraenkel:2021ijv,Feldman:2019upn,Murciano:2019wdl,Murciano:2020lqq,Murciano:2022lsw,Calabrese:2020tci,Turkeshi:2020yxd,Parez:2020vsp,Parez:2021pgq,Ares:2022hdh,Ares:2022koq,Piroli:2022ewy,Scopa:2022gfw}. These advances converge on a central insight: the equipartition of SREE emerges as a signature of conformal invariance or integrability, thereby providing a sharp criterion for universality beyond conventional entanglement measures.

In practical simulations, the explicit implementation of symmetry resolution—as manifested in tensor network algorithms such as density matrix renormalization group (DMRG), matrix product state (MPS), and multi-scale entanglement renormalization ansatz (MERA) \cite{White1992White,White:1993zza,Vidal:2007hda,Vidal:2008zz,Verstraete:2008cex}—is not merely advantageous but essential. Decomposing the wavefunction into distinct symmetry sectors $\{m\}$ dramatically reduces the resource overhead for high-precision calculations. This efficiency stems from exploiting the block-diagonal structure of symmetric density matrices, enabling studies of complex systems that would otherwise be beyond computational reach.

Recent breakthroughs in quantum simulation have transformed SREE from a purely theoretical construct into a measurable observable. Pioneering experiments with ultra-cold atoms in optical lattices \cite{Bloch:2008zzb,Kaufman:2016mif,Islam:2015mom,Lukin:2019tkq,Azses:2020tdz,Neven:2021igr,Vitale:2021lds,Rath:2022qif} have demonstrated the capability to directly probe entanglement. The development of controlled "SWAP" operations between replicated many-body states, as demonstrated with $^{87}\text{Rb}$ atomic chains, provides a concrete pathway for extracting SREE \cite{Islam:2015mom,Lukin:2019tkq,Azses:2020tdz,Azses:2022nfl,Vitale:2021lds,Neven:2021igr,Rath:2022qif}. This methodology, which bridges quantum information protocols with AMO platforms, opens unprecedented avenues for the experimental validation of symmetry-resolved quantum correlations.

Traditional approaches to SREE are primarily based on twist operator correlators in conformal field theory (CFT) \cite{Murciano:2020vgh} and minimal surface prescriptions in holography \cite{Belin:2013uta,Huang:2025lsy,zhao2021symmetry,weisenberger2021symmetry,zhao2022charged}. 
Complementary studies have also explored free bosonic systems using correlation matrix techniques \cite{Pirmoradian:2023uvt}, providing additional examples of how SREE can be computed numerically.
Recently, boundary conformal field theory (BCFT) has been employed as a systematic framework to analyze SREE for finite or continuous symmetry groups \cite{DiGiulio:2022jjd,Northe:2023khz,Kusuki:2023bsp,Choi:2024tri,Choi:2024wfm,Kusuki:2024gss}. 
The heat kernel method, originating from the seminal works of Fock, Schwinger, and later DeWitt, has become a fundamental tool in quantum field theory and mathematical physics \cite{fock1937eigenzeit,schwinger1951gauge,dewitt1965dynamical,vassilevich2003heat,Ivanov:2021fgt}. 
Given its geometric nature and natural compatibility with replica manifold techniques \cite{Solodukhin:2011gn,Dowker:1977zj,Fursaev:1994in,Mann:1997hm,Frolov:1998ea}, the heat kernel framework provides a promising basis for developing a formulation of SREE.

For a quantum field $\Phi$ governed by a Gaussian Lagrangian $\mathcal L=\Phi^\dagger D\Phi$, the effective action $W[\Phi]$ can be expressed in terms of the heat kernel $K(s,x,x^\prime)$, which solves the heat equation associated with the operator $D$:
\begin{equation}
W[\Phi] = -\frac{1}{2} \int_0^\infty \frac{ds}{s} \operatorname{Tr} K(s).
\end{equation}
In standard replica calculations \cite{Calabrese:2004eu,Calabrese:2009qy,Calabrese:2016xau}, boundary conditions on $n$-sheeted Riemann surfaces $\mathcal{R}_n$ are systematically handled using the Sommerfeld formula \cite{Fursaev:1994in,Dowker:1977zj}, which adjusts the periodicity of field configurations to account for topological defects. This approach has proven effective for neutral systems, in which the analytic structure of $\operatorname{Tr} K(s)$ remains well controlled.

However, the introduction of background $U(1)$ gauge fields $A_\mu(x)$ in charged quantum systems reveals significant limitations in the conventional Sommerfeld framework. Specifically: (i) \textit{Incomplete residue analysis}: Conventional Sommerfeld formulas account only for singularities arising from insertion functions while neglecting additional singularities induced by the non-analytic nature of heat kernels, thus leading to systematic discrepancies between residue summations and discrete Sommerfeld expansions; (ii) \textit{Breakdown of asymptotic boundary conditions}: Contour integral treatments lack rigorous justification for neglecting contributions from infinite boundaries, an approximation that fails when the chemical potential $\mu \neq 0$.

To address these challenges, this work develops a systematic approach that reconciles the analytic structure of heat kernels with the constraints imposed by a finite chemical potential.
Our approach reconstructs the heat kernel representation of the effective action through a careful reformulation of the analytic structure of Sommerfeld formulas. The main innovations are: (i) replacement of imposed periodic boundary conditions by asymptotic decay constraints; (ii) introducing a kernel function endowed with a phase factor, which systematically handles charged configurations and eliminates contributions from infinite contours. This refinement enables the first globally convergent expansion for charged heat kernels on $n$-sheeted manifolds, thus bridging the gap between discrete Sommerfeld-type expansions and continuous spectral decompositions.

We obtain explicit expressions for the $d$-dimensional charged R\'enyi entropy of scalar fields. These results not only generalize existing formulas for neutral systems, but also recover traditional Sommerfeld predictions in the $\mu\to 0$ limit, thereby validating the consistency of our approach. 

Furthermore, we verify the universality of our framework through two independent benchmarks. First, in the half-system setup, our general $d$-dimensional result reduces at d=2 and agrees exactly with the SREE obtained from twist-operator methods in CFT \cite{Murciano:2020vgh}.
Second, computations of entanglement entropy for conformal field theories on $S^1\times H ^{d-1}$ backgrounds match the analytic predictions from \cite{Belin:2013uta,Huang:2025lsy}. Together, these validations confirm that the improved heat-kernel method remains robust across both low-dimensional conformal systems and higher-dimensional geometric settings.

Finally, building on the known cMERA representation of neutral entanglement entropy \cite{Fernandez-Melgarejo:2021ymz}, we extend the surface contribution of the replicated heat kernel to Gaussian cMERA states in the presence of a chemical potential. 
The heat kernel provides a natural bridge between entanglement entropy and the cMERA formulation, through its relation to the Laplacian operator and the corresponding Green function.
In particular, we derive a modified cMERA entanglement flow equation in the charged case and demonstrate its smooth reduces to the neutral result as $\mu\to0$.

The paper is organized as follows. Section~\ref{sec:entanglement_heatkernel} reviews the definition of SREE and the representation of the heat kernel of effective actions. Section~\ref{sec:charged_sommerfeld} presents the improved Sommerfeld scheme. In Section~\ref{sec:hk_sree} this scheme is applied to compute the entropies of half system as well as on the $S^1\times H ^{d-1}$ backgrounds. Section~\ref{sec:cMERA_HK} establishes the correspondence between the surface heat kernel and Gaussian cMERA correlation functions, and derives a modified entanglement flow equation in the presence of a chemical potential. Finally, Section~\ref{sec:conclusion} summarizes the broader implications of the framework and outlines potential extensions to MERA and holographic duality.

\section{Entanglement and heat kernel}\label{sec:entanglement_heatkernel}
\subsection{Entanglement entropy and symmetry resolution}\label{sec:ee_sree}
Entanglement entropy, a fundamental measure of nonlocal quantum correlations in quantum many-body systems, has attracted considerable attention in condensed matter physics and quantum field theory over recent decades. Its mathematical foundation stems from the spectral analysis of reduced density matrices: For a pure state $\ket{\psi}$ with density matrix $\rho = \ket{\psi}\bra{\psi}$, the reduced density matrix of subsystem $A$ is defined as $\rho_A = \operatorname{Tr}_B \rho$, where $\operatorname{Tr}_B$ denotes the partial trace over the complementary subsystem $B$. The von Neumann entropy and R\'enyi entropy are correspondingly expressed as
\begin{align}
S &\equiv -\operatorname{Tr}(\rho_A \ln \rho_A), \\
S_n &\equiv \frac{1}{1-n} \ln \operatorname{Tr} \rho_A^n,
\end{align}
with the relation $S = \lim_{n \to 1} S_n$ obtained through analytic continuation from integer values of $n$ to real values. 

Within the framework of quantum field theory, the trace of the $n$-th power of the reduced density matrix could be expressed as
\begin{equation}
\operatorname{Tr}\rho^n_{A}=\frac{Z_n}{Z_1^n},
\end{equation}
where $Z_n\equiv Z[\R_n]$ denotes the partition function of the Euclidean path integral on the $n$-fold branched cover $\R_n$ of the cut spacetime manifold. The replica manifold $\R_n$ has a conical deficit with deficit angle $2\pi(1-\alpha)$ localized at the entangling surface $\Sigma$. The R\'enyi entropy can be reformulated as \cite{Calabrese:2004eu}
\begin{equation}\label{eq:REW}
S_n = \frac{1}{1-n}\left(\ln Z_n-n\ln Z_1\right)=\frac{W_n-nW_1}{n-1},
\end{equation}
with replica effective action $W_n=-\ln Z_n$. By analytic continuation from a positive integral $n$ to a non-integral $\alpha$, we express $S$ as
\begin{equation}\label{eq:EEW}
S = -\left. \frac{\partial}{\partial \alpha} \ln \frac{Z_\alpha}{Z_1^\alpha} \right|_{\alpha=1} = \left. \left( \alpha \frac{\partial}{\partial \alpha} - 1 \right) W_\alpha \right|_{\alpha=1},
\end{equation}
where the effective action $W_\alpha$ is defined via the replicated partition function:
\begin{equation}
W_\alpha \equiv -\ln Z_\alpha + \alpha \ln Z_1.
\end{equation}
Here, the term $\alpha\ln Z_1$ ensures the cancellation of boundary contributions from individual replicas. For normalized vacuum states ($Z_1=1$), this simplifies to
\begin{equation}
S = \left. \partial_\alpha W_\alpha \right|_{\alpha=1}.
\end{equation}
This framework achieves remarkable success in conformal field theory. For $(1+1)$-dimensional CFTs on an infinite line, the Calabrese-Cardy method provides analytic solutions for the R\'enyi entropy of a single interval of length $l$ through twist operator correlation functions \begin{equation}
S_n = \frac{c}{6} \left( 1 + \frac{1}{n} \right) \ln \left( \frac{l}{\epsilon} \right),\qquad S = \frac{c}{3} \ln \left( \frac{l}{\epsilon} \right),
\end{equation}
where $c$ is the central charge, $l$ is the interval length, and $\epsilon$ is the ultraviolet cutoff. For finite systems of total size $R$ with periodic boundary conditions (e.g., a circle), the entropy for half the system ($l=R/2$) becomes:
\begin{equation}
S_n = \frac{c}{12} \left( 1 + \frac{1}{n} \right) \ln \left( \frac{R}{\pi \epsilon} \right),\qquad S = \frac{c}{6} \ln \left( \frac{R}{\pi \epsilon} \right),
\end{equation}
reflecting the topological constraints imposed by the compactified spatial dimension \cite{Calabrese:2004eu,Calabrese:2009qy,Calabrese:2016xau}.

A groundbreaking advancement emerges within the AdS/CFT correspondence \cite{tHooft:1993dmi,Susskind:1994vu,Maldacena1999,Witten:1998qj,Gubser:1998bc}: The Ryu-Takayanagi formula maps holographic entanglement entropy to minimal surface areas in AdS spacetime \cite{Ryu:2006bv},
\begin{equation}
S = \frac{\text{Area}(\gamma)}{4G_N},
\end{equation}
revealing an intrinsic connection between quantum entanglement and spacetime geometry. In particular, the continuous Multi-scale Entanglement Renormalization Ansatz (cMERA) \cite{Haegeman:2011uy} -- a field-theoretic realization of tensor networks -- exhibits a profound correspondence between its renormalization group flow and AdS geometry \cite{Swingle:2009bg,Nozaki:2012zj}. This correspondence provides powerful new tools for probing the microscopic mechanisms of holography via refined entanglement entropy frameworks.

Recent extensions of this framework to SREE have yielded substantial progress. Consider a system with a conserved $U(1)$ charge generated by an operator $Q$. The Hilbert space decomposes naturally into distinct charge sectors labeled by the eigenvalues $q$. The reduced density matrix $\rho_A$ can be block-diagonalized as
\begin{align}
\rho_A = \oplus_q p_A(q) \rho_A(q), \quad p_A(q) = \operatorname{Tr}(\Pi_q \rho_A),
\end{align}
where $\Pi_q$ is the projection operator onto the charge-$q$ subspace, and $\rho_A(q)$ describes the normalized density matrix within this sector. The symmetry-resolved R\'enyi entropy for sector $q$ is then defined as
\begin{align}
S_n(q) \equiv \frac{1}{1-n} \ln \left[ \frac{Z_n(q)}{Z_1^n(q)} \right], \quad Z_n(q) = \operatorname{Tr} \left( \Pi_q \rho_A^n \right).
\end{align}
where $Z_n(q)$ are the symmetry-resolved moments.
They are related to the charged moments $Z_n(\mu)$ through a Fourier transform:
\begin{align}\label{Fourier}
Z_n(q) = \int_{0}^{2\pi} \frac{d\mu}{2\pi}\, e^{-iq\mu} Z_n(\mu),
\end{align}
so that $q$ and $\mu$ play conjugate roles as the discrete charge and its associated chemical potential.

To compute $Z_n(q)$, we introduce a fixed background $U(1)$ gauge field that couples to the conserved current \cite{Belin:2013uta} .
This background field is flat ($dA=0$) everywhere except at the entangling surface $\Sigma$,
where it carries a nontrivial Wilson line.

The charged moments $Z_n(\mu)$ are naturally expressed as a partition function in the replicated manifold $\R_n$, with the gauge field $A_\mu(x)$ introducing twisted boundary conditions across replicas. This yields:
\begin{align}\label{eq:charged_Z}
    Z_n(\mu)=\operatorname{Tr} \left[ \rho^n_A e^{i\mu Q_A} \right],
\end{align}
and the corresponding charged R\'enyi entropy reads

\begin{equation}\label{eq:Charged_RE}
S_n(\mu) = \frac{1}{n -1}\left(W_n(\mu) - n W_1(\mu)\right),
\end{equation}
where $W_n(\mu)=-\ln Z_n(\mu)$ is the charged effective action corresponding to the charged moment. 
Notice that we choose the factor $e^{i\mu Q_A}$ in Eq.~\eqref{eq:charged_Z}, instead of the factor $e^{in\mu Q_A}$ chosen in Ref.~\cite{Belin:2013uta}.

\subsection{Heat kernel representation for effective actions}
\label{sec:hk_effective_action}
The heat kernel method provides a systematic framework for computing effective actions and entanglement entropy in quantum field theory. Its particular effectiveness in curved spacetime backgrounds makes it a critical tool for evaluating entanglement entropy in nontrivial geometries, particularly the conical geometries that emerge in conformal field theory applications. The foundational analyses by Fursaev and Frolov \cite{Fursaev:1994in,Frolov:1996aj} comprehensively address its implementation on manifolds containing conical singularities.

For a bosonic field $\Phi$, the effective action $W$ under Gaussian approximation is fundamentally expressed as:
\begin{equation}\label{eq:effective_action}
    e^{-W} = (\det D)^{-1/2}, \quad  W = \frac{1}{2} \operatorname{Tr} \ln D,
\end{equation}
where $D$ is a positive-definite elliptic operator acting on the field space. In this work, unless otherwise specified, $D$ will be taken to be the Laplace operator. Assume that $D$ admits a complete orthonormal eigenbasis:
\begin{equation}
D |d_n\rangle = d_n |d_n\rangle, \quad \langle d_m|d_n\rangle = \delta_{mn}.
\end{equation}
The determinant and trace-logarithm then follow directly:
\begin{equation}
\det D = \prod_n d_n, \quad \operatorname{Tr} \ln D = \sum_n \ln d_n.
\end{equation}
Using the integral identity:
\begin{equation}
\ln d_n = -\int_0^\infty \frac{ds}{s} e^{-sd_n},
\end{equation}
the effective action transforms into:
\begin{equation}
W = -\frac{1}{2} \int_0^\infty \frac{ds}{s} \sum_n e^{-sd_n}.
\end{equation}

The corresponding heat kernel is defined as
\begin{equation}\label{eq:heat_kernel}
K(s;x,x^\prime) \equiv \langle x|e^{-sD}|x^\prime\rangle = \sum_n e^{-sd_n} \braket{x|d_n}\braket{d_n|x^\prime},
\end{equation}
satisfies the heat equation with initial condition:
\begin{equation}\label{eq:HeatKernelEq}
(\partial_s + D)K(s;x,x^\prime) = 0, \quad K(0;x,x^\prime) = \delta^{(d)}(x-x^\prime).
\end{equation}
The heat kernel trace evaluates to:
\begin{equation}\label{eq:trace_heat_kernel}
\operatorname{Tr} K(s) = \int dx\,\ K(s; x, x) = \sum_n e^{-sd_n},
\end{equation}
yielding the final expression for $W$:
\begin{equation}\label{eq:Whk}
W = -\frac{1}{2} \int_0^\infty \frac{ds}{s} \operatorname{Tr}K(s).
\end{equation}

To compute the entropies from the effective action, 
according to Eqs.~\eqref{eq:REW}, \eqref{eq:trace_heat_kernel}, and \eqref{eq:Whk}, 
it suffices to evaluate the trace of the heat kernel, rather than the full heat kernel itself.

The above formulas apply to a scalar field $\Phi$ on a general manifold with general boundary conditions. 
Different choices of manifold and boundary conditions lead to different effective actions and corresponding heat kernels, 
which will be specified in the next section.

\section{Sommerfeld formula for heat kernel}\label{sec:charged_sommerfeld}

In this section, we first review the Sommerfeld formula for constructing the heat kernel $K_\alpha$ on the replica manifold $\R_\alpha$, 
and then propose an improved Sommerfeld formula for constructing the charged heat kernel $K_{\alpha\mu}$.

For simplicity, we consider a $d$-dimensional spacetime containing a $(d-2)$-dimensional entangling surface $\Sigma$. 
We introduce coordinates 
$x=(r,\,\phi,\,\vec z)$,
where $\vec z=(z_1,\,z_2,\,\cdots,\,z_{d-2})$ denote coordinates along the $(d-2)$-dimensional hyperplane $\Sigma$, 
and $(r,\,\phi)$ are polar coordinates on the two-dimensional surface transverse to $\Sigma$. 
The angular coordinate $\phi$ has period $2\pi$.

\subsection{For neutral heat kernel}\label{sec:sommerfeld}

To compute the R\'enyi entropy via Eq.~\eqref{eq:REW}, we consider the scalar field $\Phi(x)$ defined on the replica manifold $\R_\alpha$, 
subject to the periodic boundary condition along the replica direction,
\begin{align}\label{eq:periodic_phi}
    \Phi(r,\phi+2\pi \alpha,\vec z)=\Phi(r,\phi,\vec z).
\end{align}
We denote the effective action in Eq.~\eqref{eq:effective_action} on $\R_\alpha$ by $W_\alpha$, 
and the corresponding heat kernel in Eq.~\eqref{eq:heat_kernel} by $K_\alpha(s;x,x')$. 
These quantities remain related through Eq.~\eqref{eq:Whk}. 
According to Eq.~\eqref{eq:trace_heat_kernel}, only the coincidence limit $K_\alpha(s;x,x)$ is required. 
This can be constructed from the heat kernel $K(s;x,x')$, 
which satisfies the heat equation with the initial condition in Eq.~\eqref{eq:HeatKernelEq}.

Let us consider $r=r'$ and $\vec z=\vec z'$ in $x$ and $x'$ and denote the heat kernel as
\begin{align}
    K_\alpha(s;x,x')=K_\alpha(s;\phi-\phi'),
\end{align}
where we have assumed the translational invariance along the angular $\phi$ direction and hidden the dependence on $(r,\vec z)$ in $K(s;\phi-\phi')$ following Ref.~\cite{Solodukhin:2011gn}.

Based on the periodic boundary condition imposed on $\Phi$ in Eq.~\eqref{eq:periodic_phi}, 
the heat kernel on $\mathbb{R}_\alpha$ must also satisfy $2\pi\alpha$-periodicity, namely
\begin{equation}\label{eq:PeriodicityCondition}
    K_\alpha(s; \phi - \phi') 
    = K_\alpha(s; \phi - \phi' + 2\pi \alpha).
\end{equation}
Given a heat kernel $K(s;\phi)$ that satisfies the heat equation together with the initial condition Eq.~\eqref{eq:HeatKernelEq}, 
the essence of the Sommerfeld method is to construct new solutions $K_\alpha(s;\phi)$ by linearly superposing basis kernels so as to enforce the desired boundary condition \eqref{eq:PeriodicityCondition}. 
For a conical geometry with opening angle $2\pi\alpha$, this is achieved through the discrete sum
\begin{equation}\label{eq:DiscreteSommerfeld}
    K_\alpha(s; \phi-\phi') 
    = \sum_{m\in\mathbb{Z}} 
      K\big(s; \phi-\phi' + 2\pi m\alpha\big),
\end{equation}
which automatically ensures the periodicity condition \eqref{eq:PeriodicityCondition}. 
Importantly, the kernel $K(s;\phi)$ itself is not required to satisfy the periodic boundary condition \eqref{eq:PeriodicityCondition} with $\alpha=1$.

\begin{figure}
    \centering
    \includegraphics[width=0.5\linewidth]{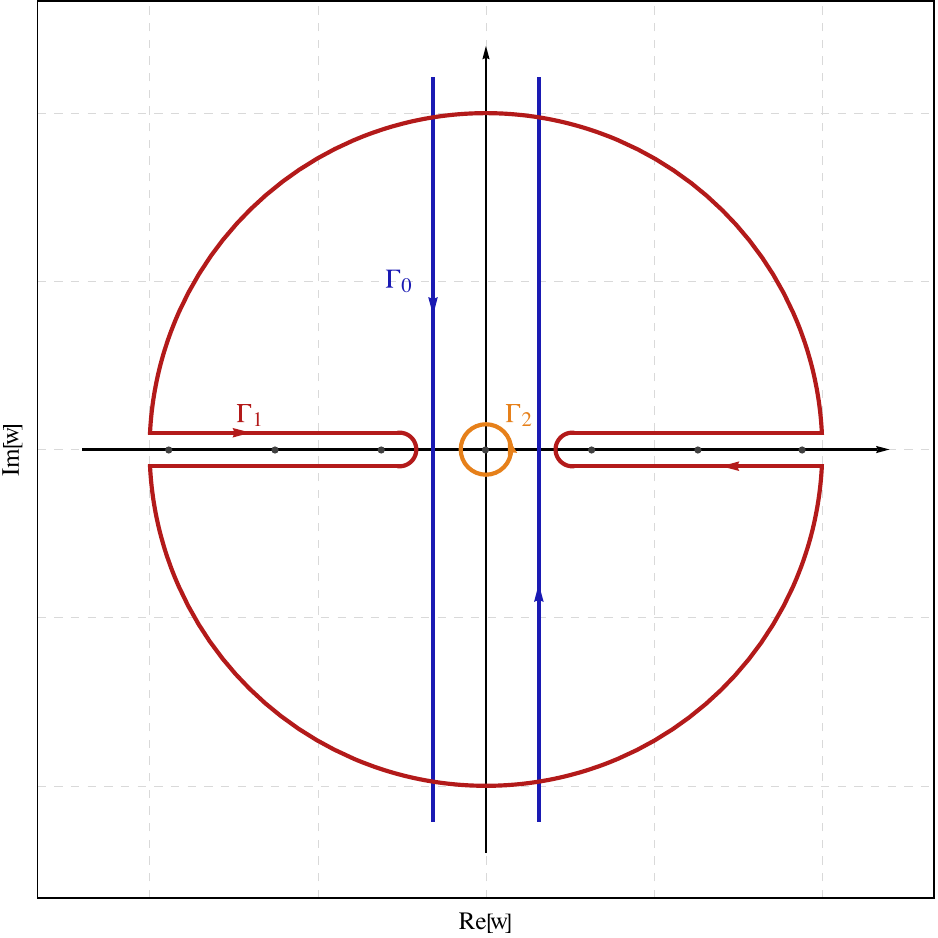}
    \caption{
  Integration contours $\Gamma_0$, $\Gamma_1$, and $\Gamma_2$ in the complex $w$-plane.
  All contours are topologically equivalent, enclosing the same set of poles except for the one at the origin.
}
    \label{fig:Contours}
\end{figure}

We can transform the summation into contour integration in the complex $w$-plane by replacing the discrete index $m$ by a set of poles of an auxiliary insertion factor, so that the discrete sum is recovered through residue calculus. One choice of the insertion factor leads to the Sommerfeld formula \cite{Dowker:1977zj, Fursaev:1994in, Solodukhin:2011gn} at $\phi=\phi'$, 
\begin{equation}\label{eq:ContinuousSommerfeld}
\begin{split}
    K_\alpha(s)\equiv&\, K_\alpha(s;0) = K(s;0) + \Delta_\alpha(s),\\
    \Delta_\alpha(s)=&\, \frac{i}{4\pi \alpha} \int_{\Gamma_0} \cot\left(\frac{w}{2\alpha}\right) K(s;w) dw,
\end{split}
\end{equation}
where the contour $\Gamma_0$ consists of two vertical segments: $(-\pi + i\infty)\to(-\pi - i\infty)$ and $(\pi - i\infty)\to(\pi + i\infty)$, as illustrated in Fig.~\ref{fig:Contours}. The real parts $\operatorname{Re}w=\pm\pi$ are chosen so that the contour encloses all poles of the integrand except the one at $w=0$.
These poles are located at $w=2\pi\alpha m (m\in\mathbb{Z})$, where $\alpha$ is the replica parameter that will be continued to values close to $1$; thus, the choice $\pm\pi$ naturally captures all nonzero poles within a $2\pi$-wide fundamental strip.
The normalization factor originates from the residue calculation:

\begin{equation}
    \mathrm{Res}\left(\frac1{2\alpha}\cot\left(\frac{w}{2\alpha}\right)f(w),\, w = 2\pi m \alpha\right) = f(2\pi m \alpha),\quad m\in\mathbb Z,
\end{equation}
for an arbitrary analytical function $f(w)$.

We notice that to let the contour integral along $\Gamma_0$ in Eq.~\eqref{eq:ContinuousSommerfeld} faithfully capture the summation $\sum_{m\in\mathbb Z}$ in \eqref{eq:DiscreteSommerfeld} except $m=0$,  one should ensure that: (1) the integrant only has poles on $w=2\pi m\alpha$; (2) the contribution from the two infinite semi-circles $\Im w\to \pm\infty$ vanish.

\subsection{For charged heat kernel}\label{sec:phase_convergence}

The conventional Sommerfeld formula suffices for neutral cases. Now we present a refined version that systematically accommodates charged fields while preserving geometric intuition.

We consider the situation where the chemical potential is purely imaginary, $\mu = i \mu_E$ with real $\mu_E$. Similarly, to compute the charged R\'enyi entropy via Eq.~\eqref{eq:Charged_RE}, we should consider the scalar field $\Phi(x)$ on the replica manifold $\R_\alpha$ but with the phase boundary condition
\begin{align}\label{eq:charged_periodic_phi}
    \Phi(r,\phi+2\pi \alpha,\vec z)=e^{i \mu_E}\Phi(r,\phi,\vec z).
\end{align}
We denote its effective action with Eq.~\eqref{eq:charged_periodic_phi} as $W_\alpha(\mu)$ and its heat kernel as $K_{\alpha\mu}(s)$. Similarly, the heat kernel satisfies the phase boundary conditions
\begin{equation}\label{eq:pbc}
    K_{\alpha\mu}(s;\phi-\phi'+ 2\pi \alpha) = e^{i\mu_E} K_{\alpha\mu}(s;\phi-\phi'),
\end{equation}
leading to the discrete spectral representation
\begin{equation}\label{eq:charged_heat_kernel}
    K_{\alpha\mu}(s; \phi-\phi') = \sum_{m \in \mathbb Z} e^{-im\mu_E} K(s; \phi - \phi' + 2\pi m\alpha).
\end{equation}
where $K(s,\phi)$ is the same as that in last subsection.

We further notice the periodicity $2\pi$ of the heat kernel on $\mu_E$
\begin{align}\label{eq:perodicity}
    K_{\alpha\mu}(s;\phi-\phi')=K_{\alpha(\mu+i2\pi)}(s;\phi-\phi').
\end{align}
So,  without loss of generality, we will first consider the case of $\mu_E\in[0,2\pi)$ and then extend the final analytical result to the complete domain by periodicity Eq.\,\eqref{eq:perodicity}.

Here, we also rewrite the summation \eqref{eq:charged_heat_kernel} into a contour integral by introducing the improved Sommerfeld formula with $\mu_E\in[0,2\pi)$ at $\phi=\phi'$,
\begin{equation}\label{eq:ImprovedSommerfeld}
\begin{split}
    K_{\alpha\mu}(s) \equiv&\, K_{\alpha\mu}(s;0)= K(s;0) + \Delta_{\alpha\mu}(s)\\
    \Delta_{\alpha\mu}(s) =&\,\frac{i}{2\pi} \int_{\Gamma_1} \Xi(w) K(s; w) dw =\frac{i}{2\pi} \int_{\Gamma_2} \Xi(w) K(s; w) dw,
\end{split}
\end{equation}
with the contours $\Gamma_1,\,\Gamma_2$ in Fig.~\ref{fig:Contours} and the designed factor
\begin{equation}
    \Xi(w) \equiv \frac{i}{\alpha} e^{-i \frac{w}{2\pi\alpha}\mu_E}\frac{1}{1 - e^{-iw/\alpha}}
\end{equation}
in order to pickup the sum of residues according to 
\begin{equation}
    \mathrm{Res}\left(\Xi(w)f(w),\ w = 2\pi m\alpha\right) 
    = e^{-i m\mu_E}f(2\pi m\alpha),\quad m\in \mathbb Z
\end{equation}
and eliminate the contribution along both the infinite semi-circles $\mathrm{Im}\, \omega\to \pm\infty$ in the contour $\Gamma_1$ by utilizing the asymptotic behaviors
\begin{align}
    &\Xi(w) \sim \exp\kd{-\kc{1-\frac{\mu_E}{2\pi}}\frac{\Im w}{\alpha}}\to0,\quad \Im w\to +\infty.  \\
    &\Xi(w) \sim \exp\kd{-\frac{\mu_E}{2\pi}\frac{\Im w}{\alpha}}\to0,\quad \Im w\to -\infty. 
\end{align}
By assuming no other poles in the integrand in Eq.~\eqref{eq:ImprovedSommerfeld} beside $w=2\pi m\alpha$ for $m\in\mathbb Z$, we can deform the contour along $\Gamma_1\to\Gamma_0\to \Gamma_2$ and then reduce the contour integral to the residue at $w=0$. Therefore, we require that the heat kernel $K(s;w)$ have no poles in the complex $w$-plane except at the origin and grow slowly than exponentially as $\Im w \to \pm \infty$.

\section{Entropies for free scalar field}\label{sec:hk_sree}

\subsection{Heat kernel for free scalar field}
\label{sec:polar_consistency}
We now verify the applicability of the improved Sommerfeld formula to $d$-dimensional free bosonic fields.  
To this end, we first solve the heat kernel and its trace in polar coordinates before applying the replica trick to evaluate the entanglement entropy.
Consider a $d$-dimensional spacetime containing a $(d-2)$-dimensional entangling surface. 
The solution to the heat equation Eq.~\eqref{eq:HeatKernelEq} obtained by Fourier transformation reads
\begin{equation}\label{eq:FourierHeatKernel}
    K(s; x, x') = \frac{1}{(2\pi)^d} \int d^dp \; e^{ip_\mu(x^\mu - x'^\mu)} e^{-s p^2}.
\end{equation}

To evaluate the $K(s;w)$ in  Eqs.~\eqref{eq:ContinuousSommerfeld}\eqref{eq:ImprovedSommerfeld}, we set $r=r'$, $\phi=\phi'+w$, and $\vec z = \vec z'$ in Eq.~\eqref{eq:FourierHeatKernel}. With this choice, the distance becomes $|x-x'|=2r\sin(w/2)$.  
Let $p=|p_\mu|$ and $\theta$ be the angle between $p_\mu$ and $(x-x')_\mu$, then
\begin{equation}
    p_\mu (x^\mu - x'^{\mu}) = 2pr \sin\!\left(w/2\right) \cos\theta.
\end{equation}

As mentioned before, $K(s;w)$ is not required to obey the periodic boundary condition, so it is no longer necessary to preserve the exact trigonometric form of $\sin(w/2)$.
Accordingly, we may expand it around $w=0$ :
\begin{equation}
    \sin(w/2)=\frac{w}{2} - \frac{w^3}{48} + O(w^4),
\end{equation}
and retain the two leading orders. This truncation removes the zeros at $\omega = 2\pi n$ for all $n\in\mathbb{Z}$ with $n\neq 0$. 
As we will show, the trace of the heat kernel, $\Tr K(s;w)$, becomes analytic on the real axis except at the origin, 
thereby satisfying the assumptions required for the improved Sommerfeld formula Eq.~\eqref{eq:ImprovedSommerfeld}.
Substituting this expansion gives
\begin{equation}
    p_\mu (x^\mu - x'^{\mu})
    \simeq 2pr\left(\frac{w}{2} - \frac{w^3}{48}\right)\cos\theta,
\end{equation}
which will be sufficient for the subsequent small-$w$ analysis.

The integration measure for the $d$-dimensional momentum $p_\mu$ in spherical coordinates reads:
\begin{equation}
    \int d^dp = \Omega_{d-2} \int p^{d-1} dp \int \sin^{d-2} \theta d\theta,
\end{equation}
where $\Omega_{d-2} = 2\pi^{(d-1)/2}/\Gamma[(d-1)/2]$. Substituting into Eq.~\eqref{eq:FourierHeatKernel} yields:
\begin{align}\label{eq:TraceKernelPolar}
    K(s; w, r)
= \frac{\Omega_{d-2}}{(2\pi)^d}
  \int_0^\infty \! p^{d-1} \, dp
  \int_0^\pi \! \sin^{d-2}\!\theta \,
  e^{i\, 2 p r \left(\frac{w}{2} - \frac{w^3}{48}\right) \cos\theta}
  e^{- s p^2}
  d\theta,
\end{align}
where we have shown the $r$-dependence of $K(s;w)$ explicitly. 
The angular integral can be simplified using the Bessel function identity:
\begin{equation}
    \int_0^\pi \sin^\nu \theta e^{ix \cos \theta} d\theta = \sqrt{\pi}\, {\Gamma\left(\frac{\nu + 1}{2}\right)}
    \left(\frac{2}{x}\right)^{\nu/2} J_{\nu/2}(x), \label{eq:BesselIdentity}
\end{equation}
where $J_{\nu/2}(x)$ is the Bessel function of the first kind, we obtain
\begin{align}
K(s; w, r) = \frac{2}{(4\pi)^{d/2}}
  \frac{1}{\bigl[r\,(\frac{w}{2}-\frac{w^3}{48})\bigr]^{\frac{d-2}{2}}}
  \int_0^\infty dp\, p^{d/2}\,
  J_{\frac{d-2}{2}}\Bigl(2pr\,(\frac{w}{2}-\frac{w^3}{48})\Bigr)\, e^{-s p^2}.
\end{align}

The full trace over configuration space requires integration over $r$, $\phi$, and transverse coordinates $\vec z$:
\begin{align}
\operatorname{Tr}K(s;w) &= \int d^{d-2}z \int_0^{2\pi}d\phi \int_0^\infty rdr \, K(s;w,r) \\
&=\frac{s}{(4\pi s)^{d/2}}\cdot\pi \alpha A(\Sigma)\cdot \left(\frac{4}{w^2}+\frac{1}{3}\right) \label{eq:SimpTraceKernelPolar},
\end{align}
where $A(\Sigma) = \int d^{d-2}z$ is the entangling surface area. This result agrees with the known expressions in \cite{Solodukhin:2011gn}.

\subsection{Entanglement entropy of half system}
\label{sec:multi_riemann}\label{sec:ee_hk}

Substituting our trace expression Eq.~\eqref{eq:SimpTraceKernelPolar} into the trace of improved Sommerfeld formula in Eq.~\eqref{eq:ImprovedSommerfeld} with $\mu_E=0$ produces
\begin{align}
    \operatorname{Tr} \Delta_\alpha=&\,
      \frac{1}{2\pi \alpha} \int_{\Gamma_2} \frac{1}{1 - e^{-iw/\alpha}} \operatorname{Tr}K(s; w) dw \label{eq:SommerfeldTrace}\\
     =&\, \frac{s}{(4\pi s)^{d/2}} \frac{A(\Sigma)}{2}\int_{\Gamma_2} \frac{1}{1 - e^{-i \frac{w}{\alpha} }}  \left(\frac{4}{w^2}+\frac{1}{3}\right) dw
     = \frac{A(\Sigma)\alpha C_2(\alpha)}{2(4\pi s)^{(d-2)/2}}, \label{eq:TrDelta}
\end{align}
where
\begin{align}
    C_2(\alpha) \equiv \frac{1}{4\pi\alpha} \int_{\Gamma_2} \frac{1}{1 - e^{-i \frac{w}{\alpha} \, }}  \left(\frac{4}{w^2}+\frac{1}{3}\right)dw.
\end{align}
As the contour $\Gamma_2$ encloses all poles except the origin, only the residue at $w=0$ contributes:
\begin{align}
    \mathrm{Res}\left(\frac{1}{1 - e^{-i \frac{w}{\alpha} \, }}  \left(\frac{4}{w^2}+\frac{1}{3}\right), w = 0\right) &= \frac{i \left(\alpha ^2-1\right)}{3 \alpha },
\end{align}
yielding
\begin{align}\label{c2}
    C_2(\alpha) = \frac{1}{6}\left(\frac{1}{ \alpha^2}-1\right).
\end{align}
Combining the above results, the heat kernel on a multi-sheeted Riemann surface takes the form
\begin{align}
\operatorname{Tr} K_\alpha(s) = \frac{V\alpha}{(4\pi s)^{d/2}}  + \frac{A(\Sigma)\alpha C_{2}(\alpha)}{2(4\pi s)^{(d-2)/2}}.
\end{align}

The effective action can be evaluated as
\begin{align}
W_\alpha = -\frac{1}{2}\int_{\epsilon^2}^\infty \frac{ds}{s} \operatorname{Tr} K_\alpha(s)
= -\frac{1}{2}\int_{\epsilon^2}^\infty \frac{ds}{s} \left[\frac{V\alpha}{(4\pi s)^{d/2}}  + \frac{A(\Sigma)\alpha C_{2}(\alpha)}{2(4\pi s)^{(d-2)/2}}\right]. \label{WaKa}
\end{align}
Using Eqs.~\eqref{eq:REW} and \eqref{eq:EEW}, we obtain
\begin{align}
S_\alpha &=
\frac{A(\Sigma)}
{12(d-2)(4\pi)^{(d-2)/2}\epsilon^{d-2}}
\left(1+\frac{1}{\alpha}\right), \\
S &=
\frac{A(\Sigma)}
{6(d-2)(4\pi)^{(d-2)/2}\epsilon^{d-2}}.
\end{align}
This consistency in neutral systems provides the foundation for extending the formalism to charged fields.

\subsection{SREE of half system}\label{sec:sree_scalar}

We compute the SREE of a free complex scalar field in flat $d$-dimensional space with a planar $(d-2)$-dimensional entangling surface $\Sigma$. 

Utilizing the heat kernel trace expression derived in Eq.~(\ref{eq:SimpTraceKernelPolar}), the trace of the improved Sommerfeld formula Eq.~\eqref{eq:ImprovedSommerfeld} become
\begin{align}
    \operatorname{Tr} \Delta_{\alpha\mu} = \frac{1}{2\pi\alpha} \int_{\Gamma_2} e^{-i\frac{w}{2\pi\alpha}\mu_E}\frac{1}{1 - e^{-i \frac{w}{\alpha} \, }} \operatorname{Tr}K(s; w) dw    =\frac{A(\Sigma)\,\alpha\, C_2(\alpha,\mu_E)}{2(4\pi s)^{(d-2)/2}},
\end{align}
where 
\begin{align}
    C_2(\alpha,\mu_E) \equiv \frac{1}{4\pi\alpha} \int_{\Gamma_2} e^{-i\frac{w}{2\pi\alpha}\mu_E}\frac{1}{1 - e^{-i \frac{w}{\alpha} \, }} \left(\frac{4}{w^2}+\frac{1}{3}\right)dw.
\end{align}
The residue calculation at $w=0$ yields the essential singularity structure:
\begin{align}
  & \mathrm{Res}\left(e^{-i\frac{w}{2\pi\alpha}\mu_E}\frac{1}{1 - e^{-i \frac{w}{\alpha} \, }} \left(\frac{4}{w^2}+\frac{1}{3}\right), w = 0\right)\nonumber \\ 
   &= -\frac{i \mu_E }{\pi  \alpha }+\frac{i \mu_E ^2}{2 \pi ^2 \alpha  }+\frac{i }{3 \alpha }-\frac{i \alpha }{3 }.
\end{align}
Consequently, the coefficient function takes the form:
\begin{align}\label{eq:c2_charged}
C_2(\alpha,\mu_E)= \frac{1}{6}\left(\frac{1}{ \alpha^2}-1\right)+\frac{\mu_E ^2}{4 \pi ^2 \alpha ^2}-\frac{\mu_E }{2 \pi  \alpha ^2}.
\end{align}

For charged systems, the charged R\'enyi entropy is expressed in terms of the effective action $W_n(\mu_E)$ as (\ref{eq:Charged_RE}), where the effective action can be computed via the heat kernel trace as
\begin{align}
    W_n(\mu_E) &= -\frac{1}{2}\int_{\epsilon^2}^\infty \frac{ds}{s} \operatorname{Tr} K_{n\mu}(s)\\
    &= -\frac{n \left[(d - 2) V+2 \pi d \, \epsilon^2 \, A(\Sigma) \, C_2(n,\mu_E) \right]}{d (d - 2)(4\pi)^{d/2} \, \epsilon^{d}}.
\end{align}
Thus, the R\'enyi entropy is
\begin{align}
    S_n(\mu_E) = -\frac{A(\Sigma)}{2(d - 2)(4\pi)^{(d - 2)/2} \epsilon^{d - 2}} \frac{n\left[C_2(n,\mu_E) - n C_2(1,\mu_E)\right]}{n - 1}.
\end{align}
Taking the analytic continuation $n \to 1$, we obtain the $\mu_E$-dependent entanglement entropy:
\begin{align}\label{smuhk}
    S(\mu_E) = \frac{A(\Sigma)}{(d - 2)(4\pi)^{(d - 2)/2} \epsilon^{d - 2}} \left(\frac{1}{6} - \frac{\mu_E}{2\pi} + \frac{\mu_E^2}{4 \pi^2}\right).
\end{align}

For the special case of dimensions $d=2$, the power-law divergence becomes logarithmic. Introducing an infrared length scale $L$, corresponding to the size of the entangling interval, the entropy takes the form
\begin{align}
    S(\mu_E) = \frac{A(\Sigma)}{2} \ln \left(\frac{L}{\epsilon}\right) \left(\frac{\mu_E}{\pi} - \frac{\mu_E^2}{2 \pi^2} - \frac{1}{3}\right).
\end{align}

The expressions above correspond to the charged R\'enyi entropy evaluated in the grand canonical ensemble, where the chemical potential $\mu_E$ is fixed. To obtain the symmetry-resolved entanglement entropy, one must instead work in the fixed charge sector. This is achieved by performing a Fourier transform of the charged moments with respect to the chemical potential.

The charged moments are related to the effective action through $Z_n(\mu_E) = e^{-W_n(\mu_E)}$. For the special case of $d = 2$, using the effective action derived above, we obtain
\begin{align}
    \ln Z_\alpha(\mu_E)=\ln\left(\frac{L}{\epsilon}\right)\,\left[\frac{1}{3}\left(\frac{1}{ \alpha}-\alpha\right)+\frac{\mu_E ^2}{4 \pi ^2 \alpha }-\frac{\mu_E }{2 \pi  \alpha }\right].
\end{align}
To project onto a fixed charge sector labeled by $q$ , we perform the Fourier transform defined in (\ref{Fourier}). 
\begin{align}
    Z_\alpha(q)=Z_\alpha(\mu_E=0)\frac{\alpha \ln\left(\frac{L}{\epsilon}\right) }{n^2\pi^2q^2+\ln^2\left(\frac{L}{\epsilon}\right)}.
\end{align}
Applying the standard replica construction and taking the von Neumann limit, we obtain
\begin{align}
    S(q)\simeq S-\ln\ln\left(\frac{L}{\epsilon}\right)-1,
\end{align}
in agreement with known conformal field theory results \cite{Murciano:2020vgh}.

\subsection{SREE for $S^1\times H^{d-1}$ background}\label{sec:sree_s1Hd}

We consider a $d$-dimensional conformal field theory in flat space in its vacuum state. The subsystem $A$ is a spherical region of radius $R$ at fixed time. Our goal is to compute the SREE associated with the reduced density matrix $\rho_A$.

By the Casini–Huerta–Myers mapping \cite{Casini:2011kv,Belin:2013uta,Huang:2025lsy}, the domain of dependence of $A$ is mapped to $S^1 \times H^{d-1}$. 
Although this geometry is often used in holographic contexts, our analysis is entirely field-theoretic and is based on heat-kernel methods.
Under this mapping, the reduced density matrix is unitarily equivalent to a thermal density matrix,
\begin{align}
    \rho_A = U^{-1} e^{-\beta_0 H} U ,
\qquad
\beta_0 = 2\pi R ,
\end{align}
where $H$ is the Hamiltonian generating translations along the Euclidean time circle of $S^1 \times H^{d-1}$.
The Rényi entropy follows from
\begin{align}
    \mathrm{Tr}\,\rho_A^n
=
\frac{Z(\beta=n\beta_0)}{Z(\beta_0)^n} ,
\end{align}
where $Z(\beta)$ is the thermal partition function on S$^1_\beta \times H^{d-1}$.

For symmetry resolution, we consider the charged moments
$\mathrm{Tr}\!\left(\rho_A^n e^{i\mu Q_A}\right)$.
Under the same mapping, this becomes the grand-canonical partition function on $S^1_\beta \times H^{d-1}$ with inverse temperature $\beta=n\beta_0$ and chemical potential $\mu$.

The product geometry of $S^1 \times H^{d-1}$ enables factorization of the heat kernel:
\begin{align}
    K_{S^1 \times H^{d-1}}(s; x, y) = K_{S^1}(s; x_1, y_1) K_{H^{d-1}}(s; x_2, \dots, x_d, y_2, \dots, y_d),
\end{align}
where $K_{S^1}$ and $K_{H^{d-1}}$ are the heat kernels on the circle $S^1$ and hyperbolic space $H^{d-1}$, respectively.

The corresponding effective action on $S^1 \times H^{d-1}$ can be expressed as an integral of the heat kernel:
\begin{align}
    W = - \frac{1}{2}  \int \frac{ds}{s} \operatorname{Tr}\,2\,e^{-m_cs}\,K_{S^1 \times H^{d-1}}( s),
\end{align}
where the factor of 2 in front of the  heat kernel appears because a complex scalar field has twice the degrees of freedom of a real  scalar field. In addition, the curvature of $H^{d-1}$ induces an effective conformal mass term, which we denote by
\begin{align}
    	m_c = \xi R = -\left(\frac{d-2}{2}\right)^2 .
\end{align}
where $\xi=\frac{d-2}{4(d-1)}$ is the conformal coupling in $d$ dimensions and
$R=-(d-1)(d-2)$ is the Ricci scalar of $H^{d-1}$.
Here $m_c$ denotes the conformal mass term induced by the curvature of $H^{d-1}$ for a conformally coupled scalar field.

The charged system imposes phase-twisted boundary conditions on the heat kernel, consistent with the convention introduced in (\ref{eq:pbc}):
\begin{align}\label{eq:SHbc}
    K_{S^1 \times H^{d-1}}(s; x_1+2\pi m \alpha, \dots, x_d, y_1, \dots, y_d) =  e^{i m  \mu_E} K_{S^1 \times H^{d-1}}(s; x, y),
\end{align}
where $\mu_E$ represents the purely imaginary chemical potential. 
According to our definition of $Z_n(\mu)$ in Eq.~\eqref{eq:charged_Z}, we notice that the choice of boundary condition in Eq.~\eqref{eq:SHbc} here is different from that in Ref.~\cite{Belin:2013uta}.
The heat kernel $K_{S^1}(s; x_1, y_1)$ can be expressed as the heat kernel on ${\mathbb{R}}^1$ :
\begin{align}
    K_{\mathbb{R}^1}(s;x_1,y_1)=\frac{1}{(4\pi s)^{1/2}}e^{\frac{-(y_1-x_1)^2}{4s}},
\end{align}
using the discrete Sommerfeld formula:
\begin{align}
    K_{S^1}(s;x_1, y_1) = \frac{1}{(4\pi s)^{1/2}} \sum_{m \in \mathbb{Z}} e^{-\frac{(y_1 - x_1 + 2\pi \alpha m)^2}{4s}} e^{-im \mu_E} .
\end{align}

For the d-dimensional theory at inverse temperature $\beta = 2\pi \alpha$, the finite-$\mu_E$ effective action takes the form:
\begin{align}
    W_\alpha(\mu_E) = - \frac{1}{2} (2\pi\alpha) V_{H^{d-1}} \int \frac{ds}{s} \frac{2}{(4\pi s)^{1/2}} e^{-m_c s} \sum_{m \in \mathbb{Z}} e^{-\frac{\pi^2 \alpha^2 m^2}{s}} e^{-im \mu_E} \operatorname{Tr} K_{H^{d-1}}(s),
\end{align}
where $y_1-x_1=0$ was assumed for the trace. The renormalized effective action eliminates the divergent contribution associated with the $m=0$ mode, yielding the following:
\begin{align}
\hat W_\alpha(\mu_E)
=
\sum_{m\in\mathbb Z,\,m\neq0}
e^{-im\mu_E}
G_\alpha(\mu_E,m),
\end{align}
where
\begin{align}
G_\alpha(\mu_E,m)
=
-
\frac{1}{2}
(2\pi\alpha)
V_{H^{d-1}}
\int
\frac{ds}{s}
\frac{2}{(4\pi s)^{1/2}}
e^{-m_c s}
e^{-\frac{\pi^2\alpha^2 m^2}{s}}
\operatorname{Tr} K_{H^{d-1}}(s).
\end{align}
The summation over $m$ can be evaluated using the Sommerfeld formula. Introducing the kernel $\frac{1}{1 - e^{-2 \pi i z \, }}$, whose poles are located at $\{z=m \in \mathbb{Z}\}$. By inserting this function to adjust the position of the poles, we can transform the sum into a contour integral:
\begin{align}
    \hat{W}_\alpha(\mu_E) =  \int_{\Gamma_2} \frac{e^{-i z\mu_E}}{1 - e^{-2 \pi i z \, }}G_\alpha(\mu_E,m)dz.
\end{align}

In this section, the Sommerfeld formula is applied directly to the effective action rather than to the trace of the heat kernel. The two are connected through the integral Eq.~\eqref{eq:Whk}. The difference in methodology here lies only in the order of integration. Applying the residue theorem, the effective action reduces to the contribution from the pole at the origin,
\begin{align}
    \hat W_\alpha(\mu_E)
=
-2\pi i\,\;
\mathrm{Res}
\left(
\frac{e^{-i m\mu_E}}{1-e^{-2\pi i m\,}}\,
G_\alpha(\mu_E,m)\,,\,m=0
\right).
\end{align}

To evaluate the effective action explicitly, we require the heat kernel on $H^{d-1}$, whose general form is known \cite{Grigoryan:1998,Lewkowycz:2012qr}. For odd-dimensional hyperbolic spaces, it admits a closed analytic expression, while for even-dimensional cases it takes an integral form and typically requires numerical evaluation.

For $d-1=2m+1$, the heat kernel on $H^{d-1}$ can be written as
\begin{align}
    K_{H^{2m+1}}(\rho,s)
=
\frac{(-1)^m}{2^m \pi^m}\,
\frac{1}{\sqrt{4\pi s}}\,
\Bigl(\frac{1}{\sinh\rho}\frac{\partial}{\partial\rho}\Bigr)^{m}
e^{-m^2 s-\frac{\rho^2}{4s}}.
\end{align}
where $\rho$ is the geodesic distance on $H^{d-1}$. The trace of the heat kernel is obtained by taking the coincident-point limit,
\begin{align}
    \operatorname{Tr} K_{H^{2m+1}}(s)\equiv
\lim_{\rho\to 0} K_{H^{2m+1}}(\rho,s).
\end{align}
Evaluating this limit explicitly for the first few values of $m$, we obtain:
\begin{align}
    \operatorname{Tr}K_{H^{1}}(s)
&=
\frac{1}{2\sqrt{\pi}\, s^{1/2}},\\
\operatorname{Tr}K_{H^{3}}(s)
&=
\frac{e^{-s}}{8\pi^{3/2}\, s^{3/2}},\\
\operatorname{Tr}K_{H^{5}}(s)
&=
\frac{e^{-4s}}{96\pi^{5/2}\, s^{5/2}}
\left(3+2s\right),\\
\operatorname{Tr}K_{H^{7}}(s)
&=
\frac{e^{-9s}}{1920\pi^{7/2}\, s^{7/2}}
\left(15+30s+16s^2\right).\\
&\cdots\nonumber
\end{align}

Using the heat kernel results given above, the effective action $\hat W_\alpha(\mu_E)$ and entropy $S(\mu_E)$ can be evaluated explicitly. We present the results for the first few even dimensions.

For $d=2$
\begin{align}
    \hat W_\alpha(\mu_E)=
-\frac{V_{H^1}}{12\pi^2\alpha}
\Bigl(2\pi^2+3\mu_E^2-6\pi\mu_E\Bigr),
\end{align}
\begin{align}
    S(\mu_E)=
-\frac{V}{6\pi^2}
\Bigl(
2\pi^2
+3\mu_E^2
-6\pi\mu_E
\Bigr).
\end{align}

For $d=4$
\begin{align}
    \hat W_\alpha(\mu_E)=
\frac{V_{H^3}}{2880\pi^5\alpha^3}
\Bigl(-8\pi^4+60\pi^2\mu_E^2+15\mu_E^4-60\pi\mu_E^3\Bigr),
\end{align}
\begin{align}
    S(\mu_E)=
-\frac{V}{720\pi^5}
\Bigl(
8\pi^4
-60\pi^2\mu_E^2
-15\mu_E^4
+60\pi\mu_E^3
\Bigr).
\end{align}

For $d=6$
\begin{align}
    \hat W_\alpha(\mu_E)=
    &-\frac{V_{H^5}}{241920\pi^8\alpha^5}
    \Bigl[
    16\pi^6(2+7\alpha^2)
    -168\pi^4(1+5\alpha^2)\mu_E^2 \notag\\
    &-210\pi^2(-1+\alpha^2)\mu_E^4
    +21\mu_E^6 \notag+840\pi^3\alpha^2\mu_E^3
    -126\pi\mu_E^5
    \Bigr],
\end{align}
\begin{align}
    S(\mu_E)=
-\frac{V}{120960\pi^8}
\Bigl(
320\pi^6
-2184\pi^4\mu_E^2
+1680\pi^3\mu_E^3
+210\pi^2\mu_E^4
-378\pi\mu_E^5
+63\mu_E^6
\Bigr).
\end{align}

These calculations demonstrate the systematic application of our improved Sommerfeld formalism to curved backgrounds. The same computational procedure applies uniformly across different spacetime dimensions, providing a consistent framework for evaluating the effective action and entropy at finite chemical potential.

For $d=2$ and $d=4$, our results agree exactly with those reported in \cite{Huang:2025lsy}, providing a check of our method. For higher dimensions such as $d=6$, our formalism applies straightforwardly and yields explicit expressions.

To obtain the symmetry-resolved entanglement entropy, we implement the same Fourier projection onto fixed charge sectors as in the previous model. We now apply this procedure to the $S^1 \times H^{d-1}$ setup and explicitly work out the $d=2$ example. The charged partition function is given by
\begin{align}
    Z_\alpha(\mu_E)
=
\exp\bigl(-\hat W_\alpha(\mu_E)\bigr),
\end{align}
where
\begin{align}
\hat W_\alpha(\mu_E)=
-\frac{V_{H^1}}{12\pi^2\alpha}
\Bigl(2\pi^2+3\mu_E^2-6\pi\mu_E\Bigr).
\end{align}
Projecting onto the fixed charge sector labeled by integer $q$, we obtain
\begin{align}
    Z_\alpha(q)
=
\int_0^{2\pi}
\frac{d\mu_E}{2\pi}\,
e^{-\hat W_\alpha(\mu_E)}
e^{-i q \mu_E}.
\end{align}
This integral can be evaluated exactly. Expanding the result in the large-volume limit $V_{H^1}\to\infty$, we find
\begin{align}
    Z_\alpha(q)
=
\exp\left(\frac{V_{H^1}}{6\alpha}\right)
\, \left[
\frac{2\alpha}{V_{H^1}}
+
\frac{4\alpha^2}{V_{H^1}^2}
+
\frac{8\alpha^3(3-\pi^2 q^2)}{V_{H^1}^3}
+
\frac{48\alpha^4(5-2\pi^2 q^2)}{V_{H^1}^4}
+
\mathcal{O}(V_{H^1}^{-5})
\right].
\end{align}
The symmetry-resolved R\'enyi entropy can be obtained from the replica construction. Taking the von Neumann limit $\alpha \to 1$, we obtain the symmetry-resolved entanglement entropy
\begin{align}
    S(q)=
\frac{V_{H^1}}{3}
-
\ln V_{H^1}
+
\ln 2
-
1
+
\frac{4\pi^2 q^2 - 10}{V_{H^1}^2}
+
\frac{120\pi^2 q^2 -296}{3 V_{H^1}^3}
+
\mathcal{O}(V_{H^1}^{-4}).
\end{align}

\section{cMERA and heat kernel approach to EE}\label{sec:cMERA_HK}
\subsection{From MERA to cMERA}\label{sec:mera_cmera}
Having established the heat kernel approach for SREE, we now explore its deep connections to cMERA -- a tensor network framework that naturally encodes scale-dependent entanglement structures. While SREE can also be studied using approaches such as twist-operator methods, the heat kernel formulation is particularly suitable for establishing a direct connection with cMERA. This is because the heat kernel is determined by the spectrum of the Laplacian operator, whose inverse defines the Green function, and the Green function can be explicitly computed within the cMERA framework in terms of the variational entangler. In the half-space geometry, the entanglement entropy can be expressed directly in terms of the surface contribution of the heat kernel, without taking derivatives of the effective action. As a result, obtaining an analytic representation of the charged heat kernel that allows one to isolate its surface contribution is a necessary step for extending the cMERA construction to SREE, which constitutes one of the primary motivations for developing the present framework.

The MERA provides a tensor network representation of quantum states that incorporates entanglement renormalization \cite{Vidal:2007hda}.
Its architecture realizes a real-space renormalization group (RG) flow through alternating layers of disentanglers ($U$) and isometries ($W$), which respectively remove short-range correlations and perform coarse-graining.
Through this layered structure, the MERA efficiently captures the logarithmic scaling of entanglement in critical systems.

For completeness, the discrete construction using $U$ and $W$ is summarized in Appendix~\ref{app:mera}.
In the present section we focus on its continuum analog, where the generators $K(u)$ and $L$ play corresponding roles in quantum field theory.

The cMERA provides a non-perturbative framework for constructing quantum states in continuum field theories that encode scale-dependent entanglement structures \cite{Haegeman:2011uy}. Unlike its discrete counterpart MERA, which operates on lattice systems, cMERA introduces a continuous flow parameter $u\in(-\infty,0]$, interpolating between an infrared (IR) vacuum state $\ket{\Omega}$ as $u\to -\infty$ and an ultraviolet (UV) state $\ket{\Psi}$ at $u=0$. The evolution is governed by a unitary operator:
\begin{align}
    U(u_2,u_1)=\mathcal{P}\exp\left({-i}\int_{u_1}^{u_2}D(u)du\right),\quad D(u)=K(u)+L,
\end{align}
where $L$ generates scale transformations, $K(u)$ introduces entanglement across momentum scales, and $\mathcal{P}$ denotes $u$-ordering. 

Let $a(k)$ and $a^\dagger(k)$ denote annihilation and creation operators in $d$-dimensional momentum space, satisfying:
\begin{align}
    [a(k),a^\dagger(k^\prime)]=\delta^d(k-k^\prime),
\end{align}
Under a scale transformation $k\to e^u k$, these operators transform as:
\begin{align}
    a(k)\to e^{\frac{d}{2}u}a(e^u k),\quad a^\dagger(k)\to e^{\frac{d}{2}u}a^\dagger (e^u k).
\end{align}
For an infinitesimal transformation $e^u=1+\epsilon$, the variation $\delta a(k)=\epsilon(k\partial_k+d/2)a(k)$ defines the generator $L$ via:
\begin{align}\label{[L,a]}
    -i[L,a(k)]=\left(k\frac{\partial}{\partial k}+\frac{d}{2}\right)a(k).
\end{align}
Integrating Eq.~\eqref{[L,a]} yields:
\begin{align}
    L=\int d^dk\, \left[ a^\dagger(k)\left(k\frac{\partial}{\partial k}+\frac{d}{2}\right)a(k)+\mathrm{h.c.}\right].
\end{align}

A momentum-dependent Bogoliubov rotation mixes $a(k)$ and $a^\dagger(-k)$:
\begin{align}
    a(k)\to\cosh f(k,u)a(k)+\sinh f(k,u)a^{\dagger}(-k).
\end{align}
 For infinitesimal $\delta u$, $\delta a(k)=g(k,u)\delta u\,a^{\dagger}(-k)$ , the generator $K(u)$ must satisfy:
\begin{align}
    -i[K(u),a(k)]=g(k,u)a^\dagger(-k).
\end{align}
Then we identify:
\begin{align}
    K(u)=\frac{i}{2}\int d^dk\,g(u,k)\,[a^\dagger(k)a^\dagger(-k)-a(-k)a(k)],
\end{align}
where $K(u)$ introduces correlations between momentum modes, and $g(u,k)$ parametrizes entanglement strength.

Expressing $L$ and $K(u)$ in terms of $\phi(k) = \frac{1}{\sqrt{2\omega_k}}(a(k) + a^\dagger(-k))$ and $\pi(k) = -i\sqrt{\frac{\omega_k}{2}}(a(k) - a^\dagger(-k))$:
\begin{align}
L=\frac{1}{2}\int dk [\pi(-k)(k\partial_k+\frac{1}{2})\phi(k)+h.c.], \\
K(u)=\frac{1}{2}\int dkg(k,u)[\pi(-k)\phi(k)+h.c.].
\end{align}
Acting on the field operator $\phi(x)$, we derive:
\begin{align}
-i[D(u),\phi(k)]&= -\left(k\partial_k + \frac{d}{2}+g(k,u)\right)\phi(k), \\
-i[D(u),\pi(k)]&= -\left(k\partial_k + \frac{d}{2}-g(k,u)\right)\pi(k), 
\end{align}
\begin{align}
    U^{-1}(0,u)\phi(k)U(0,u)&=e^{-f(k,u)}e^{-\frac{d}{2}u}\phi(e^{-u}k),\\
    U^{-1}(0,u)\pi(k)U(0,u)&=e^{f(k,u)}e^{-\frac{d}{2}u}\pi(e^{-u}k),
\end{align}
where
\begin{align}
    f(k,u)=\int_{0}^ug(ke^{-s},s)ds.
\end{align}

The entanglement entropy is generally determined by both the geometry of the entangling surface $\Sigma$ and its embedding in ambient spacetime. However, in certain simple cases, the entropy depends solely on the intrinsic geometry of $\Sigma$.

\subsection{SREE in Gaussian cMERA}\label{sec:sree_cmera}
As established in Section~\ref{sec:ee_hk} for neutral systems, the half-space entanglement entropy depends solely on the intrinsic geometry of the entangling surface $\Sigma$. This geometric insight can be naturally extended to Gaussian cMERA states.

Using the effective action $W(\alpha)$, the entropy can be written as
\begin{align}
S = \left(\alpha \partial_\alpha - 1 \right) W(\alpha) \big|_{\alpha=1}, \quad W(\alpha) = \sum_{i=0}^\infty w_i (1-\alpha)^i,
\end{align}
so that 
\begin{align}
    S = -(w_0 + w_1).
\end{align} 

For a $d$-dimensional planar surface of area $V$ in a neutral theory, the heat kernel trace exhibits the following universal form:
\begin{equation}
\operatorname{Tr} K(s) = \frac{V}{(4\pi s)^{d/2}}.
\label{eq:neutral_trace}
\end{equation}
As established in Eq.~(\ref{WaKa}), when extended to the replicated manifold via the Sommerfeld formula, this transforms to:
\begin{align}
    \operatorname{Tr} K_{\alpha}(s) &= \alpha \operatorname{Tr} K(s)+\operatorname{Tr}\Delta_\alpha\\
    &=\frac{1}{(4\pi s)^{d/2}}  V\cdot\alpha + \frac{1}{(4\pi s)^{(d-2)/2}} A(\Sigma)\cdot\frac{\alpha C_{2}(\alpha)}{2}\\
     &=\alpha \operatorname{Tr} K(s)+\frac{\alpha C_{2}(\alpha)}{2}\operatorname{Tr}K_\Sigma(s),
\label{eq:replicated_trace}
\end{align}
where the surface term \(\operatorname{Tr}\Delta_\alpha\) originates from the $(d-2)$-dimensional entangling surface $\Sigma$ \cite{Callan:1994py}, \(K_\Sigma\) denotes the trace of the heat kernel on \(\Sigma\), and the geometric coefficient $C_{2}(\alpha)$ is derived in Eq.~(\ref{c2}):
\begin{equation}
\alpha C_{2}(\alpha) = \frac{1}{6}\left(\frac{1}{\alpha} - \alpha\right)\sim\frac{1}{3}(1-\alpha)+O\left((\alpha -1)^2\right).
\label{eq:c2_alpha}
\end{equation}
Expanding \(\operatorname{Tr}\Delta_\alpha\) near \(\alpha=1\) gives the leading contribution relevant for entanglement:
\begin{align}
    \operatorname{Tr} \Delta_{\alpha} = \operatorname{Tr}K_\Sigma(s)\cdot  \frac{1}{6}\cdot (1-\alpha).
\end{align}

Integrating over the proper time $s$, 
\begin{align}
    w_0=0,\quad
    w_1=-\frac{1}{2}\int \frac{ds}{s}\operatorname{Tr}K_\Sigma(s)\cdot  \frac{1}{6},
\end{align}

The formalism naturally extends to charged quantum fields through the modified Sommerfeld prescription developed in Section~\ref{sec:charged_sommerfeld}. For a system with chemical potential $\mu=i\mu_E$, the Sommerfeld-corrected heat kernel trace reads
\begin{align}
\operatorname{Tr} K_{\alpha\mu}(s) = \alpha \operatorname{Tr} K(s) + \operatorname{Tr}\Delta_{\alpha\mu} 
= \alpha \operatorname{Tr} K(s) + \frac{\alpha C_2(\alpha,\mu_E)}{2} \operatorname{Tr} K_\Sigma(s),
\end{align}
where the coefficient function $C_2(\alpha,\mu_E)$ from Eq.~(\ref{eq:c2_charged}) takes the form
\begin{align}
C_2(\alpha,\mu_E) = \frac{1}{6}\left(\frac{1}{ \alpha^2}-1\right)+\frac{\mu_E ^2}{4 \pi ^2 \alpha ^2}-\frac{\mu_E }{2 \pi  \alpha ^2}.
\end{align}
Expanding near \(\alpha = 1\) gives the correction in the effective action:
\begin{align}
    \alpha\left[\frac{1}{6}\left(\frac{1}{ \alpha^2}-1\right)+ \frac{ \mu_E^2}{4\pi^2}- \frac{|\mu_E| }{2\pi \alpha}\right]
\sim
\frac{\mu_E ^2-2 \pi  \mu_E }{4 \pi ^2}- \left(\frac{\mu_E }{2 \pi }-\frac{\mu_E ^2}{4 \pi ^2}-\frac{1}{3}\right)(1-\alpha )+O\left((1- \alpha)^2\right).
\end{align}
This yields
\begin{align}
    w_1+w_0=-\frac{1}{2}\int\frac{ds}{s}\operatorname{Tr}K_\Sigma(s)\cdot\left(\frac{1}{6} - \frac{\mu_E}{2\pi} + \frac{\mu_E^2}{4 \pi^2}\right).
\end{align}
the entanglement entropy is expressed as
\begin{align}
    S=\frac{1}{12}\int_{\epsilon^2}^\infty\frac{ds}{s}\operatorname{Tr}K_{\Sigma}(s)=-\frac{1}{12}\ln\det(-\Delta(\Sigma)),
\end{align}
and the charged entanglement entropy becomes
\begin{align}
S(\mu_E) &= \left(\frac{1}{12} - \frac{\mu_E}{4\pi} + \frac{\mu_E^2}{8 \pi^2}\right) \int_{\epsilon^2}^{\infty} \frac{ds}{s} \operatorname{Tr} K_\Sigma(s) \nonumber\\
&= -\left(\frac{1}{12} - \frac{\mu_E}{4\pi} + \frac{\mu_E^2}{8 \pi^2}\right) \ln \det(-\Delta(\Sigma)).
\end{align}
where $\Delta(\Sigma)$ is the Laplacian operator defined on the $(d-2)$-dimensional surface $\Sigma$. The entropy can then be expressed in terms of the Green function as
\begin{align}\label{SGF}
    S=
\frac{A(\Sigma)}{6}
\int d^{d-1}k_\Sigma
\ln\bra{\Psi_\Lambda}\phi(k_\Sigma)\phi(-k_\Sigma)\ket{\Psi_\Lambda}+C.
\end{align}

This result enables us to establish a direct relation between the entanglement entropy and cMERA.
To this end, consider a Gaussian state $\ket\Psi$ annihilated by the operator
\begin{align}
a_k = \sqrt{\frac{\alpha(k)}{2}}  \phi(k) + i \sqrt{\frac{1}{2\alpha(k)}} \pi(k),
\end{align}
such that $a_k\ket\Psi=0$. This condition implies a precise relation between the field operators $\phi(k)$ and $\pi(k)$ when acting on $\ket\Psi$:
\begin{align}
\phi(k) \ket{\Psi} = -i \frac{1}{\alpha(k)} \pi(k) \ket{\Psi}, \quad
\bra{\Psi} \phi(k) = i \frac{1}{\alpha(k)} \bra{\Psi} \pi(k).
\end{align}

\noindent The commutation relation $[\phi(k),\pi(p)]=i\delta(k+p)$ governs the evaluation of expectation values. Substituting the above relations into this structure, one finds:
\begin{align}
\braket{\pi(k)\pi(p)} &= \frac{\alpha(k)}{2}\delta(k+p), &
\braket{\phi(k)\phi(p)} &= \frac{1}{2\alpha(k)}\delta(k+p), &
\braket{\phi(k)\pi(p)} &= \frac{i}{2}\delta(k+p).
\end{align}
In the expression Eq.~\eqref{SGF}, the cMERA UV state $\ket{\Psi_\Lambda}$ is obtained by evolving the IR state $\ket\Omega$, i.e., $\ket{\Psi_\Lambda}=U(0,-\infty)\ket\Omega$. For the IR state, take $\alpha(k)=M$ \cite{Hu:2017rsp,Franco-Rubio:2017tkt}, Using the evolution
\begin{align}
U^{-1}(0,-\infty) \phi(k) U(0,-\infty) = e^{-f(k,-\infty)} \phi(k),
\end{align}
the two-point function in the UV state reads
\begin{align}
\bra{\Psi_\Lambda}\phi(k_\Sigma)\phi(-k_\Sigma)\ket{\Psi_\Lambda} = e^{-2f(k_\Sigma,-\infty)} \frac{1}{2M} \delta(0).
\end{align}

Thus, the entropy can be expressed in terms of the functions in cMERA as
\begin{align}
    S=
-\frac{A(\Sigma)}{3}
\int d^{d-1}k
f(k,\infty)+C^\prime.
\end{align}
Introducing \(g(k,u) = g(u) \Gamma(k/\Lambda)\) and defining the cMERA weight function 
\begin{align}
    \Sigma(u) = \int d^{d-1} k\, k^{d-2} \Gamma(ke^{-u}/\Lambda),
\end{align}
we arrive at the differential form \cite{Fernandez-Melgarejo:2021ymz}
\begin{align}
    \frac{dS}{du}=-\frac{A(\Sigma)}{3}\cdot g(u)\cdot\Sigma(u).
\end{align}

Correspondingly, the cMERA charged entanglement flow equation becomes
\begin{align}\label{cMERA-SREE}
\frac{dS(\mu
_E)}{du} = -\left(\frac{1}{3} - \frac{\mu_E}{\pi} + \frac{\mu_E^2}{2 \pi^2}\right) A(\Sigma) g(u) \Sigma(u),
\end{align}
which generalizes the neutral case and explicitly shows the suppression of entropy growth due to the chemical potential. This result demonstrates how the background charge deforms the entanglement structure at each renormalization group scale, which is directly reflected in the modified profile of the entangler function $g(u;\mu_E)$. The expression provides a direct bridge between charged field theory and geometric entropy within the holographically motivated Gaussian cMERA framework, explicitly showing how the chemical potential suppresses entropy growth at all scales.

\section{Conclusions}\label{sec:conclusion}
The improved Sommerfeld formula developed in this work resolves challenges in computing entanglement entropy for charged quantum fields. By incorporating a phase factor into the kernel function, we systematically handle the singularities induced by chemical potentials while preserving the geometric intuition of the replica trick. This modification ensures rigorous handling of contour integrals and residue contributions, enabling precise calculations of symmetry-resolved entropies in both flat and curved spacetimes.

Our key contributions include three main results. 
First, our half-system results hold in arbitrary dimensions and reproduce the known $d=2$ logarithmic scaling \cite{Calabrese:2004eu,Calabrese:2009qy,Calabrese:2016xau}.
On $S^1\times H^{d-1}$, we obtain analytic results in even dimensions, recovering known lower-dimensional holographic formulas \cite{Huang:2025lsy} and extending them to higher dimensions.
Second, independent validations against twist operator correlators in $(1+1)D$ CFT \cite{Murciano:2020vgh} and effective action calculations in higher-dimensional AdS/CFT setups \cite{Belin:2013uta,Huang:2025lsy} confirm the robustness of our method. 
Third, extending the known cMERA representation of neutral entanglement entropy \cite{Fernandez-Melgarejo:2021ymz}, we establish a heat-kernel/cMERA correspondence for charged systems.
We derive a modified entanglement flow equation in the presence of a finite chemical potential and show that the Gaussian cMERA formalism naturally captures the microscopic mechanism of symmetry resolution across renormalization scales.

Thus, the improved heat kernel method provides a powerful geometric tool for probing symmetry-resolved entanglement, unifying the treatment of charged and neutral sectors, and offering a versatile approach applicable to both conformal field theories and their holographic duals.

Although the heat-kernel method successfully establishes a link between cMERA and SREE – achieved here through calculation correlation functions within the cMERA framework and their subsequent use in the heat-kernel expression for entropy – this connection remains indirect. It relies on the heat kernel as an intermediary computational tool that interprets cMERA-generated data into entanglement measures \cite{Fernandez-Melgarejo:2021ymz}. 

A profound challenge and a central direction for future work is therefore to develop a direct cMERA formulation of SREE. This would move beyond the current dependence on auxiliary field-theoretic machinery, constructing instead a symmetry-adapted cMERA where $U(1)$ charge conservation is manifestly encoded at every scale $u$ in the renormalization group flow. Key objectives include developing techniques to extract charged moments $Z_n(q)$ or the resolved density matrix with symmetry $\rho_A(q)$ directly from the cMERA state, potentially by leveraging the interaction between the disentangler $K(u)$ and the emergent geometry. This would bypass the need for both the heat kernel and replica trick in the SREE calculations for these states.

Such a direct cMERA-SREE framework could be utilized to study how symmetry resolution evolves under the renormalization group. Key questions include: Does equipartition \cite{Xavier:2018kqb} emerge universally along the flow towards critical points? Moreover, how does the fine-grained structure of symmetry sectors in the boundary cMERA state relate to specific geometric features (e.g., charged minimal surfaces, flux threads, or defect structures) in the bulk gravitational dual? A direct cMERA approach to SREE promises a more intrinsic understanding of the holographic dictionary for symmetry-resolved entanglement. Establishing such connections would provide a richer microscopic picture of how gauge symmetries and entanglement intertwine in quantum gravity \cite{Swingle:2009bg}.

The pursuit of this direct connection represents more than a technical refinement -- it seeks to uncover the intrinsic entanglement structures woven into the renormalization group flow itself. Success would solidify cMERA not just as an efficient variational tool but as a fundamental framework capable of directly revealing how global symmetries organize and constrain quantum entanglement across scales – a question of deep significance for quantum field theory, quantum information, and our understanding of holography.

Finally, our calculation of SREE within the cMERA formalism provides valuable insight for discrete MERA tensor networks. Equation~\eqref{cMERA-SREE} implies that the $U(1)$ charge effectively reduces the bond dimension of the charge-sector tensors along the minimal surface in MERA.

\acknowledgments
We thank Giuseppe Di Giulio, Aleksandr Ivanov, Mohammad Reza Tanhayi, and Suting Zhao for helpful discussions. ZYX acknowledges funding from the DFG through the Collaborative Research Center SFB 1170 “ToCoTronics” (Project ID 258499086 – SFB 1170) and support from the Berlin Quantum Initiative.

\appendix
\section{Discrete MERA construction}\label{app:mera}

For completeness, we summarize here the discrete formulation of the multi-scale entanglement renormalization ansatz (MERA), which provides the operational intuition behind the continuum version (cMERA) discussed in Sec.~\ref{sec:cMERA_HK}. 
The discrete MERA is built from alternating layers of disentanglers $W$ and isometries $U$, which act locally to remove short-range correlations and perform coarse-graining in the Hilbert space. 
Their structure follows from the basic principles of bipartite entanglement and Schmidt decomposition, as reviewed below.

For a Hilbert space decomposition $\mathcal{H}_A \otimes \mathcal{H}_B$ with $\dim \mathcal{H}_A = \dim \mathcal{H}_B = d$, any pure state admits the canonical Schmidt form:
\begin{equation}
|\psi\rangle = \sum_{\alpha=1}^m \lambda_\alpha |\alpha\rangle_A \otimes |\alpha\rangle_B,\quad
\lambda_1 \geq \lambda_2 \geq \cdots > 0,\quad
\sum_{\alpha=1}^m \lambda_\alpha^2 = 1,
\tag{A.1}
\end{equation}
where the Schmidt rank $m \leq d$ quantifies bipartite entanglement.  
A local unitary operator $U: \mathbb{C}^d \otimes \mathbb{C}^d \to \mathbb{C}^d \otimes \mathbb{C}^d$ can be chosen to rotate local bases and concentrate entanglement into a smaller number of Schmidt modes:
\begin{equation}
U|\psi\rangle = \sum_{\beta=1}^{\tilde{m}} \tilde{\lambda}_\beta |\tilde{\beta}\rangle_A \otimes |\tilde{\beta}\rangle_B,\quad
\tilde{m} \ll m.
\tag{A.2}
\end{equation}
This step corresponds to the action of a disentangler in the MERA layer.

The subsequent isometry $W: \mathbb{C}^d \otimes \mathbb{C}^d \to \mathbb{C}^\chi$ ($\chi \leq d^2$) maps the two-site Hilbert space to a reduced effective space:
\begin{equation}
W = \sum_{k=1}^\chi |k\rangle_C \,(\langle \tilde{k}|_A \otimes \langle \tilde{k}|_B),
\tag{A.3}
\end{equation}
yielding a compressed state,
\begin{equation}
WU|\psi\rangle = \sum_{k=1}^\chi \tilde{\lambda}_k |k\rangle_C,
\tag{A.4}
\end{equation}
with truncation error 
\begin{equation}
\epsilon = \sum_{k=\chi+1}^{d^2} \tilde{\lambda}_k^2.
\tag{A.5}
\end{equation}

The full MERA tensor network arises from iterative application of the disentangler–isometry pair across multiple layers:
\begin{equation}
|\psi^{(\ell+1)}\rangle = W^{(\ell)}U^{(\ell)}|\psi^{(\ell)}\rangle,\quad
\ell = 0,\ldots,L-1,
\tag{A.6}
\end{equation}
where $\mathcal{H}^{(\ell)}$ denotes the Hilbert space at depth $\ell$.  
The sequence of bond dimensions $\{\chi_\ell\}$ decreases monotonically,
\begin{equation}
\chi_0 \geq \chi_1 \geq \cdots \geq \chi_L \geq 1,
\tag{A.7}
\end{equation}
encoding the gradual loss of microscopic degrees of freedom along the renormalization direction.

\bibliographystyle{JHEP}
\bibliography{ref}

\providecommand{\href}[2]{#2}\begingroup\raggedright\begin{thebibliography}{100}

\bibitem{Eisert:2008ur}
J.~Eisert, M.~Cramer and M.B.~Plenio, \emph{{Area laws for the entanglement entropy - a review}}, \href{https://doi.org/10.1103/RevModPhys.82.277}{\emph{Rev. Mod. Phys.} {\bfseries 82} (2010) 277} [\href{https://arxiv.org/abs/0808.3773}{{\ttfamily 0808.3773}}].

\bibitem{Eisert:2006kue}
J.~Eisert, \emph{{Entanglement in quantum information theory}},  other thesis, 10, 2006, [\href{https://arxiv.org/abs/quant-ph/0610253}{{\ttfamily quant-ph/0610253}}].

\bibitem{Horodecki:2009zz}
R.~Horodecki, P.~Horodecki, M.~Horodecki and K.~Horodecki, \emph{{Quantum entanglement}}, \href{https://doi.org/10.1103/RevModPhys.81.865}{\emph{Rev. Mod. Phys.} {\bfseries 81} (2009) 865} [\href{https://arxiv.org/abs/quant-ph/0702225}{{\ttfamily quant-ph/0702225}}].

\bibitem{Levin:2006zz}
M.~Levin and X.-G.~Wen, \emph{{Detecting Topological Order in a Ground State Wave Function}}, \href{https://doi.org/10.1103/PhysRevLett.96.110405}{\emph{Phys. Rev. Lett.} {\bfseries 96} (2006) 110405} [\href{https://arxiv.org/abs/cond-mat/0510613}{{\ttfamily cond-mat/0510613}}].

\bibitem{Holzhey:1994we}
C.~Holzhey, F.~Larsen and F.~Wilczek, \emph{{Geometric and renormalized entropy in conformal field theory}}, \href{https://doi.org/10.1016/0550-3213(94)90402-2}{\emph{Nucl. Phys. B} {\bfseries 424} (1994) 443} [\href{https://arxiv.org/abs/hep-th/9403108}{{\ttfamily hep-th/9403108}}].

\bibitem{Calabrese:2009qy}
P.~Calabrese and J.~Cardy, \emph{{Entanglement entropy and conformal field theory}}, \href{https://doi.org/10.1088/1751-8113/42/50/504005}{\emph{J. Phys. A} {\bfseries 42} (2009) 504005} [\href{https://arxiv.org/abs/0905.4013}{{\ttfamily 0905.4013}}].

\bibitem{Calabrese:2004eu}
P.~Calabrese and J.L.~Cardy, \emph{{Entanglement entropy and quantum field theory}}, \href{https://doi.org/10.1088/1742-5468/2004/06/P06002}{\emph{J. Stat. Mech.} {\bfseries 0406} (2004) P06002} [\href{https://arxiv.org/abs/hep-th/0405152}{{\ttfamily hep-th/0405152}}].

\bibitem{Calabrese:2016xau}
P.~Calabrese and J.~Cardy, \emph{{Quantum quenches in 1 + 1 dimensional conformal field theories}}, \href{https://doi.org/10.1088/1742-5468/2016/06/064003}{\emph{J. Stat. Mech.} {\bfseries 1606} (2016) 064003} [\href{https://arxiv.org/abs/1603.02889}{{\ttfamily 1603.02889}}].

\bibitem{Calabrese:2009kka}
P.~Calabrese, J.~Cardy and B.~Doyon, \emph{{Entanglement entropy in extended quantum systems}}, \href{https://doi.org/10.1088/1751-8121/42/50/500301}{\emph{J. Phys. A A} {\bfseries 42} (2009) 500301}.

\bibitem{Cardy:2007mb}
J.L.~Cardy, O.A.~Castro-Alvaredo and B.~Doyon, \emph{{Form factors of branch-point twist fields in quantum integrable models and entanglement entropy}}, \href{https://doi.org/10.1007/s10955-007-9422-x}{\emph{J. Statist. Phys.} {\bfseries 130} (2008) 129} [\href{https://arxiv.org/abs/0706.3384}{{\ttfamily 0706.3384}}].

\bibitem{Casini:2005rm}
H.~Casini, C.D.~Fosco and M.~Huerta, \emph{{Entanglement and alpha entropies for a massive Dirac field in two dimensions}}, \href{https://doi.org/10.1088/1742-5468/2005/07/P07007}{\emph{J. Stat. Mech.} {\bfseries 0507} (2005) P07007} [\href{https://arxiv.org/abs/cond-mat/0505563}{{\ttfamily cond-mat/0505563}}].

\bibitem{Casini:2005zv}
H.~Casini and M.~Huerta, \emph{{Entanglement and alpha entropies for a massive scalar field in two dimensions}}, \href{https://doi.org/10.1088/1742-5468/2005/12/P12012}{\emph{J. Stat. Mech.} {\bfseries 0512} (2005) P12012} [\href{https://arxiv.org/abs/cond-mat/0511014}{{\ttfamily cond-mat/0511014}}].

\bibitem{Casini:2009sr}
H.~Casini and M.~Huerta, \emph{{Entanglement entropy in free quantum field theory}}, \href{https://doi.org/10.1088/1751-8113/42/50/504007}{\emph{J. Phys. A} {\bfseries 42} (2009) 504007} [\href{https://arxiv.org/abs/0905.2562}{{\ttfamily 0905.2562}}].

\bibitem{Casini:2011kv}
H.~Casini, M.~Huerta and R.C.~Myers, \emph{{Towards a derivation of holographic entanglement entropy}}, \href{https://doi.org/10.1007/JHEP05(2011)036}{\emph{JHEP} {\bfseries 05} (2011) 036} [\href{https://arxiv.org/abs/1102.0440}{{\ttfamily 1102.0440}}].

\bibitem{Ryu:2006bv}
S.~Ryu and T.~Takayanagi, \emph{{Holographic derivation of entanglement entropy from AdS/CFT}}, \href{https://doi.org/10.1103/PhysRevLett.96.181602}{\emph{Phys. Rev. Lett.} {\bfseries 96} (2006) 181602} [\href{https://arxiv.org/abs/hep-th/0603001}{{\ttfamily hep-th/0603001}}].

\bibitem{Miyaji:2015yva}
M.~Miyaji and T.~Takayanagi, \emph{{Surface/State Correspondence as a Generalized Holography}}, \href{https://doi.org/10.1093/ptep/ptv089}{\emph{PTEP} {\bfseries 2015} (2015) 073B03} [\href{https://arxiv.org/abs/1503.03542}{{\ttfamily 1503.03542}}].

\bibitem{Swingle:2009bg}
B.~Swingle, \emph{{Entanglement Renormalization and Holography}}, \href{https://doi.org/10.1103/PhysRevD.86.065007}{\emph{Phys. Rev. D} {\bfseries 86} (2012) 065007} [\href{https://arxiv.org/abs/0905.1317}{{\ttfamily 0905.1317}}].

\bibitem{Vidal:2007hda}
G.~Vidal, \emph{{Entanglement Renormalization}}, \href{https://doi.org/10.1103/PhysRevLett.99.220405}{\emph{Phys. Rev. Lett.} {\bfseries 99} (2007) 220405} [\href{https://arxiv.org/abs/cond-mat/0512165}{{\ttfamily cond-mat/0512165}}].

\bibitem{Vidal:2008zz}
G.~Vidal, \emph{{Class of Quantum Many-Body States That Can Be Efficiently Simulated}}, \href{https://doi.org/10.1103/PhysRevLett.101.110501}{\emph{Phys. Rev. Lett.} {\bfseries 101} (2008) 110501} [\href{https://arxiv.org/abs/quant-ph/0610099}{{\ttfamily quant-ph/0610099}}].

\bibitem{Hung:2011nu}
L.-Y.~Hung, R.C.~Myers, M.~Smolkin and A.~Yale, \emph{{Holographic Calculations of Renyi Entropy}}, \href{https://doi.org/10.1007/JHEP12(2011)047}{\emph{JHEP} {\bfseries 12} (2011) 047} [\href{https://arxiv.org/abs/1110.1084}{{\ttfamily 1110.1084}}].

\bibitem{Balasubramanian:2011wt}
V.~Balasubramanian, M.B.~McDermott and M.~Van~Raamsdonk, \emph{{Momentum-space entanglement and renormalization in quantum field theory}}, \href{https://doi.org/10.1103/PhysRevD.86.045014}{\emph{Phys. Rev. D} {\bfseries 86} (2012) 045014} [\href{https://arxiv.org/abs/1108.3568}{{\ttfamily 1108.3568}}].

\bibitem{Balasubramanian:2013lsa}
V.~Balasubramanian, B.D.~Chowdhury, B.~Czech, J.~de~Boer and M.P.~Heller, \emph{{Bulk curves from boundary data in holography}}, \href{https://doi.org/10.1103/PhysRevD.89.086004}{\emph{Phys. Rev. D} {\bfseries 89} (2014) 086004} [\href{https://arxiv.org/abs/1310.4204}{{\ttfamily 1310.4204}}].

\bibitem{Balasubramanian:2013rqa}
V.~Balasubramanian, B.~Czech, B.D.~Chowdhury and J.~de~Boer, \emph{{The entropy of a hole in spacetime}}, \href{https://doi.org/10.1007/JHEP10(2013)220}{\emph{JHEP} {\bfseries 10} (2013) 220} [\href{https://arxiv.org/abs/1305.0856}{{\ttfamily 1305.0856}}].

\bibitem{Amico:2007ag}
L.~Amico, R.~Fazio, A.~Osterloh and V.~Vedral, \emph{{Entanglement in many-body systems}}, \href{https://doi.org/10.1103/RevModPhys.80.517}{\emph{Rev. Mod. Phys.} {\bfseries 80} (2008) 517} [\href{https://arxiv.org/abs/quant-ph/0703044}{{\ttfamily quant-ph/0703044}}].

\bibitem{Nishioka:2018khk}
T.~Nishioka, \emph{{Entanglement entropy: holography and renormalization group}}, \href{https://doi.org/10.1103/RevModPhys.90.035007}{\emph{Rev. Mod. Phys.} {\bfseries 90} (2018) 035007} [\href{https://arxiv.org/abs/1801.10352}{{\ttfamily 1801.10352}}].

\bibitem{Latorre:2009zz}
J.I.~Latorre and A.~Riera, \emph{{A short review on entanglement in quantum spin systems}}, \href{https://doi.org/10.1088/1751-8113/42/50/504002}{\emph{J. Phys. A} {\bfseries 42} (2009) 504002} [\href{https://arxiv.org/abs/0906.1499}{{\ttfamily 0906.1499}}].

\bibitem{Lewkowycz:2013nqa}
A.~Lewkowycz and J.~Maldacena, \emph{{Generalized gravitational entropy}}, \href{https://doi.org/10.1007/JHEP08(2013)090}{\emph{JHEP} {\bfseries 08} (2013) 090} [\href{https://arxiv.org/abs/1304.4926}{{\ttfamily 1304.4926}}].

\bibitem{Belin:2013uta}
A.~Belin, L.-Y.~Hung, A.~Maloney, S.~Matsuura, R.C.~Myers and T.~Sierens, \emph{{Holographic Charged Renyi Entropies}}, \href{https://doi.org/10.1007/JHEP12(2013)059}{\emph{JHEP} {\bfseries 12} (2013) 059} [\href{https://arxiv.org/abs/1310.4180}{{\ttfamily 1310.4180}}].

\bibitem{Goldstein:2017bua}
M.~Goldstein and E.~Sela, \emph{{Symmetry-resolved entanglement in many-body systems}}, \href{https://doi.org/10.1103/PhysRevLett.120.200602}{\emph{Phys. Rev. Lett.} {\bfseries 120} (2018) 200602} [\href{https://arxiv.org/abs/1711.09418}{{\ttfamily 1711.09418}}].

\bibitem{Xavier:2018kqb}
J.C.~Xavier, F.C.~Alcaraz and G.~Sierra, \emph{{Equipartition of the entanglement entropy}}, \href{https://doi.org/10.1103/PhysRevB.98.041106}{\emph{Phys. Rev. B} {\bfseries 98} (2018) 041106} [\href{https://arxiv.org/abs/1804.06357}{{\ttfamily 1804.06357}}].

\bibitem{Belin:2014mva}
A.~Belin, L.-Y.~Hung, A.~Maloney and S.~Matsuura, \emph{{Charged Renyi entropies and holographic superconductors}}, \href{https://doi.org/10.1007/JHEP01(2015)059}{\emph{JHEP} {\bfseries 01} (2015) 059} [\href{https://arxiv.org/abs/1407.5630}{{\ttfamily 1407.5630}}].

\bibitem{Pastras:2014oka}
G.~Pastras and D.~Manolopoulos, \emph{{Charged R\'enyi entropies in CFTs with Einstein-Gauss-Bonnet holographic duals}}, \href{https://doi.org/10.1007/JHEP11(2014)007}{\emph{JHEP} {\bfseries 11} (2014) 007} [\href{https://arxiv.org/abs/1404.1309}{{\ttfamily 1404.1309}}].

\bibitem{Matsuura:2016qqu}
S.~Matsuura, X.~Wen, L.-Y.~Hung and S.~Ryu, \emph{{Charged Topological Entanglement Entropy}}, \href{https://doi.org/10.1103/PhysRevB.93.195113}{\emph{Phys. Rev. B} {\bfseries 93} (2016) 195113} [\href{https://arxiv.org/abs/1601.03751}{{\ttfamily 1601.03751}}].

\bibitem{Laflorencie:2014cem}
N.~Laflorencie and S.~Rachel, \emph{{Spin-resolved entanglement spectroscopy of critical spin chains and Luttinger liquids}}, \href{https://doi.org/10.1088/1742-5468/2014/11/P11013}{\emph{J. Phys. A} {\bfseries 2014} (2014) P11013}.

\bibitem{Capizzi:2020jed}
L.~Capizzi, P.~Ruggiero and P.~Calabrese, \emph{{Symmetry resolved entanglement entropy of excited states in a CFT}}, \href{https://doi.org/10.1088/1742-5468/ab96b6}{\emph{J. Stat. Mech.} {\bfseries 2007} (2020) 073101} [\href{https://arxiv.org/abs/2003.04670}{{\ttfamily 2003.04670}}].

\bibitem{Calabrese:2021wvi}
P.~Calabrese, J.~Dubail and S.~Murciano, \emph{{Symmetry-resolved entanglement entropy in Wess-Zumino-Witten models}}, \href{https://doi.org/10.1007/JHEP10(2021)067}{\emph{JHEP} {\bfseries 10} (2021) 067} [\href{https://arxiv.org/abs/2106.15946}{{\ttfamily 2106.15946}}].

\bibitem{Bonsignori:2020laa}
R.~Bonsignori and P.~Calabrese, \emph{{Boundary effects on symmetry resolved entanglement}}, \href{https://doi.org/10.1088/1751-8121/abcc3a}{\emph{J. Phys. A} {\bfseries 54} (2021) 015005} [\href{https://arxiv.org/abs/2009.08508}{{\ttfamily 2009.08508}}].

\bibitem{Estienne:2020txv}
B.~Estienne, Y.~Ikhlef and A.~Morin-Duchesne, \emph{{Finite-size corrections in critical symmetry-resolved entanglement}}, \href{https://doi.org/10.21468/SciPostPhys.10.3.054}{\emph{SciPost Phys.} {\bfseries 10} (2021) 054} [\href{https://arxiv.org/abs/2010.10515}{{\ttfamily 2010.10515}}].

\bibitem{Ma:2021zgf}
Z.~Ma, C.~Han, Y.~Meir and E.~Sela, \emph{{Symmetric inseparability and number entanglement in charge-conserving mixed states}}, \href{https://doi.org/10.1103/PhysRevA.105.042416}{\emph{Phys. Rev. A} {\bfseries 105} (2022) 042416} [\href{https://arxiv.org/abs/2110.09388}{{\ttfamily 2110.09388}}].

\bibitem{Ares:2022gjb}
F.~Ares, P.~Calabrese, G.~Di~Giulio and S.~Murciano, \emph{{Multi-charged moments of two intervals in conformal field theory}}, \href{https://doi.org/10.1007/JHEP09(2022)051}{\emph{JHEP} {\bfseries 09} (2022) 051} [\href{https://arxiv.org/abs/2206.01534}{{\ttfamily 2206.01534}}].

\bibitem{Ghasemi:2022jxg}
M.~Ghasemi, \emph{{Universal thermal corrections to symmetry-resolved entanglement entropy and full counting statistics}}, \href{https://doi.org/10.1007/JHEP05(2023)209}{\emph{JHEP} {\bfseries 05} (2023) 209} [\href{https://arxiv.org/abs/2203.06708}{{\ttfamily 2203.06708}}].

\bibitem{Murciano:2020vgh}
S.~Murciano, G.~Di~Giulio and P.~Calabrese, \emph{{Entanglement and symmetry resolution in two dimensional free quantum field theories}}, \href{https://doi.org/10.1007/JHEP08(2020)073}{\emph{JHEP} {\bfseries 08} (2020) 073} [\href{https://arxiv.org/abs/2006.09069}{{\ttfamily 2006.09069}}].

\bibitem{Horvath:2020vzs}
D.X.~Horv\'ath and P.~Calabrese, \emph{{Symmetry resolved entanglement in integrable field theories via form factor bootstrap}}, \href{https://doi.org/10.1007/JHEP11(2020)131}{\emph{JHEP} {\bfseries 11} (2020) 131} [\href{https://arxiv.org/abs/2008.08553}{{\ttfamily 2008.08553}}].

\bibitem{Horvath:2021fks}
D.X.~Horvath, L.~Capizzi and P.~Calabrese, \emph{{U(1) symmetry resolved entanglement in free 1+1 dimensional field theories via form factor bootstrap}}, \href{https://doi.org/10.1007/JHEP05(2021)197}{\emph{JHEP} {\bfseries 05} (2021) 197} [\href{https://arxiv.org/abs/2103.03197}{{\ttfamily 2103.03197}}].

\bibitem{Horvath:2021rjd}
D.X.~Horvath, P.~Calabrese and O.A.~Castro-Alvaredo, \emph{{Branch Point Twist Field Form Factors in the sine-Gordon Model II: Composite Twist Fields and Symmetry Resolved Entanglement}}, \href{https://doi.org/10.21468/SciPostPhys.12.3.088}{\emph{SciPost Phys.} {\bfseries 12} (2022) 088} [\href{https://arxiv.org/abs/2105.13982}{{\ttfamily 2105.13982}}].

\bibitem{Capizzi:2021kys}
L.~Capizzi, D.X.~Horv\'ath, P.~Calabrese and O.A.~Castro-Alvaredo, \emph{{Entanglement of the 3-state Potts model via form factor bootstrap: total and symmetry resolved entropies}}, \href{https://doi.org/10.1007/JHEP05(2022)113}{\emph{JHEP} {\bfseries 05} (2022) 113} [\href{https://arxiv.org/abs/2108.10935}{{\ttfamily 2108.10935}}].

\bibitem{Capizzi:2022jpx}
L.~Capizzi, O.A.~Castro-Alvaredo, C.~De~Fazio, M.~Mazzoni and L.~Santamar\'\i{}a-Sanz, \emph{{Symmetry resolved entanglement of excited states in quantum field theory. Part I. Free theories, twist fields and qubits}}, \href{https://doi.org/10.1007/JHEP12(2022)127}{\emph{JHEP} {\bfseries 12} (2022) 127} [\href{https://arxiv.org/abs/2203.12556}{{\ttfamily 2203.12556}}].

\bibitem{Capizzi:2022nel}
L.~Capizzi, C.~De~Fazio, M.~Mazzoni, L.~Santamar\'\i{}a-Sanz and O.A.~Castro-Alvaredo, \emph{{Symmetry resolved entanglement of excited states in quantum field theory. Part II. Numerics, interacting theories and higher dimensions}}, \href{https://doi.org/10.1007/JHEP12(2022)128}{\emph{JHEP} {\bfseries 12} (2022) 128} [\href{https://arxiv.org/abs/2206.12223}{{\ttfamily 2206.12223}}].

\bibitem{Foligno:2022ltu}
A.~Foligno, S.~Murciano and P.~Calabrese, \emph{{Entanglement resolution of free Dirac fermions on a torus}}, \href{https://doi.org/10.1007/JHEP03(2023)096}{\emph{JHEP} {\bfseries 03} (2023) 096} [\href{https://arxiv.org/abs/2212.07261}{{\ttfamily 2212.07261}}].

\bibitem{Fraenkel:2019ykl}
S.~Fraenkel and M.~Goldstein, \emph{{Symmetry resolved entanglement: Exact results in 1D and beyond}}, \href{https://doi.org/10.1088/1742-5468/ab7753}{\emph{J. Stat. Mech.} {\bfseries 2003} (2020) 033106} [\href{https://arxiv.org/abs/1910.08459}{{\ttfamily 1910.08459}}].

\bibitem{Fraenkel:2021ijv}
S.~Fraenkel and M.~Goldstein, \emph{{Entanglement measures in a nonequilibrium steady state: Exact results in one dimension}}, \href{https://doi.org/10.21468/SciPostPhys.11.4.085}{\emph{SciPost Phys.} {\bfseries 11} (2021) 085} [\href{https://arxiv.org/abs/2105.00740}{{\ttfamily 2105.00740}}].

\bibitem{Feldman:2019upn}
N.~Feldman and M.~Goldstein, \emph{{Dynamics of Charge-Resolved Entanglement after a Local Quench}}, \href{https://doi.org/10.1103/PhysRevB.100.235146}{\emph{Phys. Rev. B} {\bfseries 100} (2019) 235146} [\href{https://arxiv.org/abs/1905.10749}{{\ttfamily 1905.10749}}].

\bibitem{Murciano:2019wdl}
S.~Murciano, G.~Di~Giulio and P.~Calabrese, \emph{{Symmetry resolved entanglement in gapped integrable systems: a corner transfer matrix approach}}, \href{https://doi.org/10.21468/SciPostPhys.8.3.046}{\emph{SciPost Phys.} {\bfseries 8} (2020) 046} [\href{https://arxiv.org/abs/1911.09588}{{\ttfamily 1911.09588}}].

\bibitem{Murciano:2020lqq}
S.~Murciano, P.~Ruggiero and P.~Calabrese, \emph{{Symmetry resolved entanglement in two-dimensional systems via dimensional reduction}}, \href{https://doi.org/10.1088/1742-5468/aba1e5}{\emph{J. Stat. Mech.} {\bfseries 2008} (2020) 083102} [\href{https://arxiv.org/abs/2003.11453}{{\ttfamily 2003.11453}}].

\bibitem{Murciano:2022lsw}
S.~Murciano, P.~Calabrese and L.~Piroli, \emph{{Symmetry-resolved Page curves}}, \href{https://doi.org/10.1103/PhysRevD.106.046015}{\emph{Phys. Rev. D} {\bfseries 106} (2022) 046015} [\href{https://arxiv.org/abs/2206.05083}{{\ttfamily 2206.05083}}].

\bibitem{Calabrese:2020tci}
P.~Calabrese, M.~Collura, G.~Di~Giulio and S.~Murciano, \emph{{Full counting statistics in the gapped XXZ spin chain}}, \href{https://doi.org/10.1209/0295-5075/129/60007}{\emph{EPL} {\bfseries 129} (2020) 60007} [\href{https://arxiv.org/abs/2002.04367}{{\ttfamily 2002.04367}}].

\bibitem{Turkeshi:2020yxd}
X.~Turkeshi, P.~Ruggiero, V.~Alba and P.~Calabrese, \emph{{Entanglement equipartition in critical random spin chains}}, \href{https://doi.org/10.1103/PhysRevB.102.014455}{\emph{Phys. Rev. B} {\bfseries 102} (2020) 014455} [\href{https://arxiv.org/abs/2005.03331}{{\ttfamily 2005.03331}}].

\bibitem{Parez:2020vsp}
G.~Parez, R.~Bonsignori and P.~Calabrese, \emph{{Quasiparticle dynamics of symmetry-resolved entanglement after a quench: Examples of conformal field theories and free fermions}}, \href{https://doi.org/10.1103/PhysRevB.103.L041104}{\emph{Phys. Rev. B} {\bfseries 103} (2021) L041104} [\href{https://arxiv.org/abs/2010.09794}{{\ttfamily 2010.09794}}].

\bibitem{Parez:2021pgq}
G.~Parez, R.~Bonsignori and P.~Calabrese, \emph{{Exact quench dynamics of symmetry resolved entanglement in a free fermion chain}}, \href{https://doi.org/10.1088/1742-5468/ac21d7}{\emph{J. Stat. Mech.} {\bfseries 2109} (2021) 093102} [\href{https://arxiv.org/abs/2106.13115}{{\ttfamily 2106.13115}}].

\bibitem{Ares:2022hdh}
F.~Ares, S.~Murciano and P.~Calabrese, \emph{{Symmetry-resolved entanglement in a long-range free-fermion chain}}, \href{https://doi.org/10.1088/1742-5468/ac7644}{\emph{J. Stat. Mech.} {\bfseries 2206} (2022) 063104} [\href{https://arxiv.org/abs/2202.05874}{{\ttfamily 2202.05874}}].

\bibitem{Ares:2022koq}
F.~Ares, S.~Murciano and P.~Calabrese, \emph{{Entanglement asymmetry as a probe of symmetry breaking}}, \href{https://doi.org/10.1038/s41467-023-37747-8}{\emph{Nature Commun.} {\bfseries 14} (2023) 2036} [\href{https://arxiv.org/abs/2207.14693}{{\ttfamily 2207.14693}}].

\bibitem{Piroli:2022ewy}
L.~Piroli, E.~Vernier, M.~Collura and P.~Calabrese, \emph{{Thermodynamic symmetry resolved entanglement entropies in integrable systems}},  \href{https://arxiv.org/abs/2203.09158}{{\ttfamily 2203.09158}}.

\bibitem{Scopa:2022gfw}
S.~Scopa and D.X.~Horv\'ath, \emph{{Exact hydrodynamic description of symmetry-resolved R\'enyi entropies after a quantum quench}}, \href{https://doi.org/10.1088/1742-5468/ac85eb}{\emph{J. Stat. Mech.} {\bfseries 2208} (2022) 083104} [\href{https://arxiv.org/abs/2205.02924}{{\ttfamily 2205.02924}}].

\bibitem{White1992White}
White and R.~Steven, \emph{White, S.R.: Density matrix formulation for quantum renormalization groups. Phys. Rev. Lett. 69, 2863-2866}, {\emph{Physical Review Letters} {\bfseries 69} (1992) 2863}.

\bibitem{White:1993zza}
S.R.~White, \emph{{Density-matrix algorithms for quantum renormalization groups}}, \href{https://doi.org/10.1103/PhysRevB.48.10345}{\emph{Phys. Rev. B} {\bfseries 48} (1993) 10345}.

\bibitem{Verstraete:2008cex}
F.~Verstraete, V.~Murg and J.I.~Cirac, \emph{{Matrix product states, projected entangled pair states, and variational renormalization group methods for quantum spin systems}}, \href{https://doi.org/10.1080/14789940801912366}{\emph{Adv. Phys.} {\bfseries 57} (2008) 143}.

\bibitem{Bloch:2008zzb}
I.~Bloch, J.~Dalibard and W.~Zwerger, \emph{{Many-body physics with ultracold gases}}, \href{https://doi.org/10.1103/RevModPhys.80.885}{\emph{Rev. Mod. Phys.} {\bfseries 80} (2008) 885} [\href{https://arxiv.org/abs/0704.3011}{{\ttfamily 0704.3011}}].

\bibitem{Kaufman:2016mif}
A.M.~Kaufman, M.E.~Tai, A.~Lukin, M.~Rispoli, R.~Schittko, P.M.~Preiss et~al., \emph{{Quantum thermalization through entanglement in an isolated many-body system}}, \href{https://doi.org/10.1126/science.aaf6725}{\emph{Science} {\bfseries 353} (2016) aaf6725}.

\bibitem{Islam:2015mom}
R.~Islam, R.~Ma, P.M.~Preiss, M.E.~Tai, A.~Lukin, M.~Rispoli et~al., \emph{{Measuring entanglement entropy through the interference of quantum many-body twins}},  \href{https://arxiv.org/abs/1509.01160}{{\ttfamily 1509.01160}}.

\bibitem{Lukin:2019tkq}
A.~Lukin, M.~Rispoli, R.~Schittko, M.E.~Tai, A.M.~Kaufman, S.~Choi et~al., \emph{{Probing entanglement in a many-body\textendash{}localized system}}, \href{https://doi.org/10.1126/science.aau0818}{\emph{Science} {\bfseries 364} (2019) aau0818}.

\bibitem{Azses:2020tdz}
D.~Azses, R.~Haenel, Y.~Naveh, R.~Raussendorf, E.~Sela and E.G.~Dalla~Torre, \emph{{Identification of Symmetry-Protected Topological States on Noisy Quantum Computers}}, \href{https://doi.org/10.1103/PhysRevLett.125.120502}{\emph{Phys. Rev. Lett.} {\bfseries 125} (2020) 120502} [\href{https://arxiv.org/abs/2002.04620}{{\ttfamily 2002.04620}}].

\bibitem{Neven:2021igr}
A.~Neven et~al., \emph{{Symmetry-resolved entanglement detection using partial transpose moments}}, \href{https://doi.org/10.1038/s41534-021-00487-y}{\emph{npj Quantum Inf.} {\bfseries 7} (2021) 152} [\href{https://arxiv.org/abs/2103.07443}{{\ttfamily 2103.07443}}].

\bibitem{Vitale:2021lds}
V.~Vitale, A.~Elben, R.~Kueng, A.~Neven, J.~Carrasco, B.~Kraus et~al., \emph{{Symmetry-resolved dynamical purification in synthetic quantum matter}}, \href{https://doi.org/10.21468/SciPostPhys.12.3.106}{\emph{SciPost Phys.} {\bfseries 12} (2022) 106} [\href{https://arxiv.org/abs/2101.07814}{{\ttfamily 2101.07814}}].

\bibitem{Rath:2022qif}
A.~Rath, V.~Vitale, S.~Murciano, M.~Votto, J.~Dubail, R.~Kueng et~al., \emph{{Entanglement Barrier and its Symmetry Resolution: Theory and Experimental Observation}}, \href{https://doi.org/10.1103/PRXQuantum.4.010318}{\emph{PRX Quantum} {\bfseries 4} (2023) 010318} [\href{https://arxiv.org/abs/2209.04393}{{\ttfamily 2209.04393}}].

\bibitem{Azses:2022nfl}
D.~Azses, D.F.~Mross and E.~Sela, \emph{{Symmetry-resolved entanglement of two-dimensional symmetry-protected topological states}}, \href{https://doi.org/10.1103/PhysRevB.107.115113}{\emph{Phys. Rev. B} {\bfseries 107} (2023) 115113} [\href{https://arxiv.org/abs/2210.12750}{{\ttfamily 2210.12750}}].

\bibitem{Huang:2025lsy}
Y.~Huang and Y.~Zhou, \emph{{Symmetry-Resolved Entanglement Entropy in Higher Dimensions}},  \href{https://arxiv.org/abs/2503.09070}{{\ttfamily 2503.09070}}.

\bibitem{zhao2021symmetry}
S.~Zhao, C.~Northe and R.~Meyer, \emph{Symmetry-resolved entanglement in AdS3/CFT2 coupled to U (1) Chern-Simons theory}, {\emph{Journal of High Energy Physics} {\bfseries 2021} (2021) 1}.

\bibitem{weisenberger2021symmetry}
K.~Weisenberger, S.~Zhao, C.~Northe and R.~Meyer, \emph{Symmetry-resolved entanglement for excited states and two entangling intervals in AdS3/CFT2}, {\emph{Journal of High Energy Physics} {\bfseries 2021} (2021) 1}.

\bibitem{zhao2022charged}
S.~Zhao, C.~Northe, K.~Weisenberger and R.~Meyer, \emph{Charged moments in W3 higher spin holography}, {\emph{Journal of High Energy Physics} {\bfseries 2022} (2022) 1}.

\bibitem{Pirmoradian:2023uvt}
R.~Pirmoradian and M.R.~Tanhayi, \emph{{Symmetry-resolved entanglement entropy for local and non-local QFTs}}, \href{https://doi.org/10.1140/epjc/s10052-024-13212-8}{\emph{Eur. Phys. J. C} {\bfseries 84} (2024) 849} [\href{https://arxiv.org/abs/2311.00494}{{\ttfamily 2311.00494}}].

\bibitem{DiGiulio:2022jjd}
G.~Di~Giulio, R.~Meyer, C.~Northe, H.~Scheppach and S.~Zhao, \emph{{On the boundary conformal field theory approach to symmetry-resolved entanglement}}, \href{https://doi.org/10.21468/SciPostPhysCore.6.3.049}{\emph{SciPost Phys. Core} {\bfseries 6} (2023) 049} [\href{https://arxiv.org/abs/2212.09767}{{\ttfamily 2212.09767}}].

\bibitem{Northe:2023khz}
C.~Northe, \emph{{Entanglement Resolution with Respect to Conformal Symmetry}}, \href{https://doi.org/10.1103/PhysRevLett.131.151601}{\emph{Phys. Rev. Lett.} {\bfseries 131} (2023) 151601} [\href{https://arxiv.org/abs/2303.07724}{{\ttfamily 2303.07724}}].

\bibitem{Kusuki:2023bsp}
Y.~Kusuki, S.~Murciano, H.~Ooguri and S.~Pal, \emph{{Symmetry-resolved entanglement entropy, spectra {\&} boundary conformal field theory}}, \href{https://doi.org/10.1007/JHEP11(2023)216}{\emph{JHEP} {\bfseries 11} (2023) 216} [\href{https://arxiv.org/abs/2309.03287}{{\ttfamily 2309.03287}}].

\bibitem{Choi:2024tri}
Y.~Choi, B.C.~Rayhaun and Y.~Zheng, \emph{{Generalized Tube Algebras, Symmetry-Resolved Partition Functions, and Twisted Boundary States}},  \href{https://arxiv.org/abs/2409.02159}{{\ttfamily 2409.02159}}.

\bibitem{Choi:2024wfm}
Y.~Choi, B.C.~Rayhaun and Y.~Zheng, \emph{{Noninvertible Symmetry-Resolved Affleck-Ludwig-Cardy Formula and Entanglement Entropy from the Boundary Tube Algebra}}, \href{https://doi.org/10.1103/PhysRevLett.133.251602}{\emph{Phys. Rev. Lett.} {\bfseries 133} (2024) 251602} [\href{https://arxiv.org/abs/2409.02806}{{\ttfamily 2409.02806}}].

\bibitem{Kusuki:2024gss}
Y.~Kusuki, S.~Murciano, H.~Ooguri and S.~Pal, \emph{{Entanglement asymmetry and symmetry defects in boundary conformal field theory}}, \href{https://doi.org/10.1007/JHEP01(2025)057}{\emph{JHEP} {\bfseries 01} (2025) 057} [\href{https://arxiv.org/abs/2411.09792}{{\ttfamily 2411.09792}}].

\bibitem{fock1937eigenzeit}
V.~Fock, \emph{Die Eigenzeit in der klassischen und in der Quantenmechanik}, {\emph{Phys. Z. Sowjetunion} {\bfseries 12} (1937) 404}.

\bibitem{schwinger1951gauge}
J.~Schwinger, \emph{On gauge invariance and vacuum polarization}, {\emph{Physical Review} {\bfseries 82} (1951) 664}.

\bibitem{dewitt1965dynamical}
B.S.~DeWitt, \emph{Dynamical theory of groups and fields}, Gordon and Breach (1965).

\bibitem{vassilevich2003heat}
D.V.~Vassilevich, \emph{Heat kernel expansion: user's manual}, {\emph{Physics reports} {\bfseries 388} (2003) 279}.

\bibitem{Ivanov:2021fgt}
A.V.~Ivanov and N.V.~Kharuk, \emph{{Special functions for heat kernel expansion}}, \href{https://doi.org/10.1140/epjp/s13360-022-03176-7}{\emph{Eur. Phys. J. Plus} {\bfseries 137} (2022) 1060} [\href{https://arxiv.org/abs/2106.00294}{{\ttfamily 2106.00294}}].

\bibitem{Solodukhin:2011gn}
S.N.~Solodukhin, \emph{{Entanglement entropy of black holes}}, \href{https://doi.org/10.12942/lrr-2011-8}{\emph{Living Rev. Rel.} {\bfseries 14} (2011) 8} [\href{https://arxiv.org/abs/1104.3712}{{\ttfamily 1104.3712}}].

\bibitem{Dowker:1977zj}
J.S.~Dowker, \emph{{Quantum Field Theory on a Cone}}, \href{https://doi.org/10.1088/0305-4470/10/1/023}{\emph{J. Phys. A} {\bfseries 10} (1977) 115}.

\bibitem{Fursaev:1994in}
D.V.~Fursaev, \emph{{Spectral geometry and one loop divergences on manifolds with conical singularities}}, \href{https://doi.org/10.1016/0370-2693(94)90590-8}{\emph{Phys. Lett. B} {\bfseries 334} (1994) 53} [\href{https://arxiv.org/abs/hep-th/9405143}{{\ttfamily hep-th/9405143}}].

\bibitem{Mann:1997hm}
R.B.~Mann and S.N.~Solodukhin, \emph{{Universality of quantum entropy for extreme black holes}}, \href{https://doi.org/10.1016/S0550-3213(98)00094-7}{\emph{Nucl. Phys. B} {\bfseries 523} (1998) 293} [\href{https://arxiv.org/abs/hep-th/9709064}{{\ttfamily hep-th/9709064}}].

\bibitem{Frolov:1998ea}
V.P.~Frolov and D.~Fursaev, \emph{{Black hole entropy in induced gravity: Reduction to 2-D quantum field theory on the horizon}}, \href{https://doi.org/10.1103/PhysRevD.58.124009}{\emph{Phys. Rev. D} {\bfseries 58} (1998) 124009} [\href{https://arxiv.org/abs/hep-th/9806078}{{\ttfamily hep-th/9806078}}].

\bibitem{Fernandez-Melgarejo:2021ymz}
J.J.~Fernandez-Melgarejo and J.~Molina-Vilaplana, \emph{{On the Entanglement Entropy in Gaussian cMERA}}, \href{https://doi.org/10.1002/prop.202100093}{\emph{Fortsch. Phys.} {\bfseries 69} (2021) 2100093} [\href{https://arxiv.org/abs/2104.01551}{{\ttfamily 2104.01551}}].

\bibitem{tHooft:1993dmi}
G.~'t~Hooft, \emph{{Dimensional reduction in quantum gravity}}, {\emph{Conf. Proc. C} {\bfseries 930308} (1993) 284} [\href{https://arxiv.org/abs/gr-qc/9310026}{{\ttfamily gr-qc/9310026}}].

\bibitem{Susskind:1994vu}
L.~Susskind, \emph{{The World as a hologram}}, \href{https://doi.org/10.1063/1.531249}{\emph{J. Math. Phys.} {\bfseries 36} (1995) 6377} [\href{https://arxiv.org/abs/hep-th/9409089}{{\ttfamily hep-th/9409089}}].

\bibitem{Maldacena1999}
J.M.~Maldacena, \emph{{The Large N Limit of Superconformal Field Theories and Supergravity}}, \href{https://doi.org/10.1023/A:1026654312961}{\emph{Int. J. Theor. Phys.} {\bfseries 38} (1999) 1113} [\href{https://arxiv.org/abs/hep-th/9711200}{{\ttfamily hep-th/9711200}}].

\bibitem{Witten:1998qj}
E.~Witten, \emph{{Anti de Sitter space and holography}}, \href{https://doi.org/10.4310/ATMP.1998.v2.n2.a2}{\emph{Adv. Theor. Math. Phys.} {\bfseries 2} (1998) 253} [\href{https://arxiv.org/abs/hep-th/9802150}{{\ttfamily hep-th/9802150}}].

\bibitem{Gubser:1998bc}
S.S.~Gubser, I.R.~Klebanov and A.M.~Polyakov, \emph{{Gauge theory correlators from noncritical string theory}}, \href{https://doi.org/10.1016/S0370-2693(98)00377-3}{\emph{Phys. Lett. B} {\bfseries 428} (1998) 105} [\href{https://arxiv.org/abs/hep-th/9802109}{{\ttfamily hep-th/9802109}}].

\bibitem{Haegeman:2011uy}
J.~Haegeman, T.J.~Osborne, H.~Verschelde and F.~Verstraete, \emph{{Entanglement Renormalization for Quantum Fields in Real Space}}, \href{https://doi.org/10.1103/PhysRevLett.110.100402}{\emph{Phys. Rev. Lett.} {\bfseries 110} (2013) 100402} [\href{https://arxiv.org/abs/1102.5524}{{\ttfamily 1102.5524}}].

\bibitem{Nozaki:2012zj}
M.~Nozaki, S.~Ryu and T.~Takayanagi, \emph{{Holographic Geometry of Entanglement Renormalization in Quantum Field Theories}}, \href{https://doi.org/10.1007/JHEP10(2012)193}{\emph{JHEP} {\bfseries 10} (2012) 193} [\href{https://arxiv.org/abs/1208.3469}{{\ttfamily 1208.3469}}].

\bibitem{Frolov:1996aj}
V.P.~Frolov, D.V.~Fursaev and A.I.~Zelnikov, \emph{{Statistical origin of black hole entropy in induced gravity}}, \href{https://doi.org/10.1016/S0550-3213(96)00678-5}{\emph{Nucl. Phys. B} {\bfseries 486} (1997) 339} [\href{https://arxiv.org/abs/hep-th/9607104}{{\ttfamily hep-th/9607104}}].

\bibitem{Grigoryan:1998}
A.~Grigor'yan and M.~Noguchi, \emph{The heat kernel on hyperbolic space}, \href{https://doi.org/10.1112/S0024609398004780}{\emph{Bull. Lond. Math. Soc.} {\bfseries 30} (1998) 643}.

\bibitem{Lewkowycz:2012qr}
A.~Lewkowycz, R.C.~Myers and M.~Smolkin, \emph{{Observations on entanglement entropy in massive QFT's}}, \href{https://doi.org/10.1007/JHEP04(2013)017}{\emph{JHEP} {\bfseries 04} (2013) 017} [\href{https://arxiv.org/abs/1210.6858}{{\ttfamily 1210.6858}}].

\bibitem{Callan:1994py}
C.G.~Callan, Jr. and F.~Wilczek, \emph{{On geometric entropy}}, \href{https://doi.org/10.1016/0370-2693(94)91007-3}{\emph{Phys. Lett. B} {\bfseries 333} (1994) 55} [\href{https://arxiv.org/abs/hep-th/9401072}{{\ttfamily hep-th/9401072}}].

\bibitem{Hu:2017rsp}
Q.~Hu and G.~Vidal, \emph{{Spacetime Symmetries and Conformal Data in the Continuous Multiscale Entanglement Renormalization Ansatz}}, \href{https://doi.org/10.1103/PhysRevLett.119.010603}{\emph{Phys. Rev. Lett.} {\bfseries 119} (2017) 010603} [\href{https://arxiv.org/abs/1703.04798}{{\ttfamily 1703.04798}}].

\bibitem{Franco-Rubio:2017tkt}
A.~Franco-Rubio and G.~Vidal, \emph{{Entanglement and correlations in the continuous multi-scale entanglement renormalization ansatz}}, \href{https://doi.org/10.1007/JHEP12(2017)129}{\emph{JHEP} {\bfseries 12} (2017) 129} [\href{https://arxiv.org/abs/1706.02841}{{\ttfamily 1706.02841}}].

\end{thebibliography}\endgroup

\end{document}